\documentstyle[a4]{article}

\input{epsf}

\newcommand{\slL}{\raise.15ex\hbox{$/$}\kern-.53em\hbox{$L$}}
\newcommand{\slP}{\raise.15ex\hbox{$/$}\kern-.53em\hbox{$P$}}
\newcommand{\slR}{\raise.15ex\hbox{$/$}\kern-.53em\hbox{$R$}}
\newcommand{\slQ}{\raise.15ex\hbox{$/$}\kern-.53em\hbox{$Q$}}
\newcommand{\slK}{\raise.15ex\hbox{$/$}\kern-.53em\hbox{$K$}}

\newcommand{\be}{\begin{equation}}
\newcommand{\ee}{\end{equation}}     
\newcommand{\bea}{\begin{eqnarray}}
\newcommand{\ena}{\end{eqnarray}}

\def\build#1\over#2{\mathrel{\mathop{\kern 0pt#2}\limits_{#1}}}

\catcode`\@=11

\def\thebibliography#1{\section*{REFERENCES}\list{\arabic{enumi}.}
  {\settowidth\labelwidth{#1.}\leftmargin=1.67em
   \labelsep\leftmargin \advance\labelsep-\labelwidth
   \itemsep\z@ \parsep\z@
   \usecounter{enumi}}\def\makelabel##1{\rlap{##1}\hss}%
   \def\newblock{\hskip 0.11em plus 0.33em minus -0.07em}
   \sloppy \clubpenalty=4000 \widowpenalty=4000 \sfcode`\.=1000\relax}
   
\newcount\@tempcntc
\def\@citex[#1]#2{\if@filesw\immediate\write\@auxout{\string\citation{#2}}\fi
  \@tempcnta\z@\@tempcntb\m@ne\def\@citea{}\@cite{%
	\@ordonner{#2}%
	\@for\@citeb:=#2\do%
    {\@ifundefined{b@\@citeb}%
	{\@citeo\@tempcntb\m@ne\@citea%
        	\def\@citea{,\penalty\@m\ }{\bf ?}\@warning%
       		{Citation `\@citeb' on page \thepage \space undefined}}%
    	{\setbox\z@\hbox{\global\@tempcntc0\csname b@\@citeb\endcsname\relax}
     \ifnum\@tempcntc=\z@ \@citeo\@tempcntb\m@ne%
       \@citea\def\@citea{,\penalty\@m}%
       \hbox{\csname b@\@citeb\endcsname}%
     \else%
      \advance\@tempcntb\@ne%
      \ifnum\@tempcntb=\@tempcntc%
      \else\advance\@tempcntb\m@ne\@citeo%
      \@tempcnta\@tempcntc\@tempcntb\@tempcntc\fi\fi}}\@citeo}{#1}}%

\def\@citeo{\ifnum\@tempcnta>\@tempcntb\else\@citea
  \def\@citea{,\penalty\@m}%
  \ifnum\@tempcnta=\@tempcntb\the\@tempcnta\else
   {\advance\@tempcnta\@ne\ifnum\@tempcnta=\@tempcntb \else
\def\@citea{--}\fi
    \advance\@tempcnta\m@ne\the\@tempcnta\@citea\the\@tempcntb}\fi\fi}

\def\@toto{}
\newif\if@ordre 
\newcount\c@current
\newcount\c@last

\def\@ordonner#1{\global\c@last\m@ne%
		\global\@ordretrue%
		\@for\@toto:=#1\do%
			{\@ifundefined{b@\@toto}%
			{}%
			{\c@current\csname b@\@toto\endcsname\relax%
			\ifnum\the\c@current<\the\c@last\relax%
				{\global\@ordrefalse}\fi%
			\global\c@last\the\c@current%
			}%
			}%
		\if@ordre{}\else{\typeout{}%
			\typeout{Warning: the references are not %
			 in increasing order\on@line:}%
			\@for\@toto:=#1\do%
			{\@ifundefined{b@\@toto}%
			{}%
			\typeout{\@toto:\space \@nameuse{b@\@toto}}%
			}\typeout{}}\fi%
		}%
		
\catcode`\@=12

\catcode`\@=11

\newcount\c@subequation

\def\eqnarray{
\def\@eqnnum{{\reset@font\rm%
(\theequation-{\alph{subequation}})}}
\global\c@subequation=1\relax
\stepcounter{equation}\let\@currentlabel\theequation
\global\@eqnswtrue\m@th
\global\@eqcnt\z@\tabskip\@centering\let\\\@eqncr
$$\halign to\displaywidth\bgroup\@eqnsel\hskip\@centering
  $\displaystyle\tabskip\z@{##}$&\global\@eqcnt\@ne
  \hskip 2\arraycolsep \hfil${##}$\hfil
  &\global\@eqcnt\tw@ \hskip 2\arraycolsep $\displaystyle\tabskip\z@{##}$\hfil
   \tabskip\@centering&\llap{##}\tabskip\z@\cr}

\def\@@eqncr{\let\@tempa\relax
    \ifcase\@eqcnt \def\@tempa{& & &}\or \def\@tempa{& &}%
      \else \def\@tempa{&}\fi
     \@tempa \if@eqnsw\@eqnnum\global\advance\c@subequation by 1\relax
			\fi
     \global\@eqnswtrue\global\@eqcnt\z@\cr}

\def\endeqnarray{\@@eqncr\egroup
      \global\advance\c@equation\m@ne$$\global\@ignoretrue
	\stepcounter{equation}
	\def\@eqnnum{{\reset@font\rm (\theequation)}}}

\def\Eqnarray{
\def\@eqnnum{{\reset@font\rm (\theequation)}}
\global\c@subequation=1\relax
\stepcounter{equation}\let\@currentlabel\theequation
\global\@eqnswtrue\m@th
\global\@eqcnt\z@\tabskip\@centering\let\\\@eqncr
$$\halign to\displaywidth\bgroup\@eqnsel\hskip\@centering
  $\displaystyle\tabskip\z@{##}$&\global\@eqcnt\@ne
  \hskip 2\arraycolsep \hfil${##}$\hfil
  &\global\@eqcnt\tw@ \hskip 2\arraycolsep $\displaystyle\tabskip\z@{##}$\hfil
   \tabskip\@centering&\llap{##}\tabskip\z@\cr}

\def\endEqnarray{\@@eqncr\egroup
      \global\advance\c@equation\m@ne$$\global\@ignoretrue
	\stepcounter{equation}
	\def\@eqnnum{{\reset@font\rm (\theequation)}}}

\catcode`\@=12

\font\tenimbf=cmmib10 at 10pt
\font\sevenimbf=cmmib10 at 7pt
\font\fiveimbf=cmmib10 at 5pt
\newfam\imbf
\textfont\imbf=\tenimbf
\scriptfont\imbf=\sevenimbf
\scriptscriptfont\imbf=\fiveimbf
\def\imb{\fam\imbf\tenimbf}

\font\teneurm=eurm10
\font\seveneurm=eurm7
\font\fiveeurm=eurm5
\newfam\eurmfam
\textfont\eurmfam=\teneurm
\scriptfont\eurmfam=\seveneurm
\scriptscriptfont\eurmfam=\fiveeurm

\font\teneusm=eusm10
\font\seveneusm=eusm7
\font\fiveeusm=eusm5
\newfam\eusmfam
\textfont\eusmfam=\teneusm
\scriptfont\eusmfam=\seveneusm
\scriptscriptfont\eusmfam=\fiveeusm

\font\teneufm=eufm10
\font\seveneufm=eufm7
\font\fiveeufm=eufm5
\newfam\eufmfam
\textfont\eufmfam=\teneufm
\scriptfont\eufmfam=\seveneufm
\scriptscriptfont\eufmfam=\fiveeufm

\begin{document}
\bibliographystyle{unsrt}

\begin{titlepage}
\title{\bf{Cutting rules in the real time formalisms\\
       at finite temperature}}
\author{Fran\c cois~Gelis\footnote{email: gelis@lapp.in2p3.fr}}
\maketitle
\vskip 1cm
\begin{center}
Laboratoire de Physique Th\'eorique ENSLAPP,\\
B.~P.~110,  F--74941 Annecy-le-Vieux Cedex, France
\end{center}
\vskip 1cm
\begin{abstract}

In this paper, we review the set of rules specific to the calculation of the
imaginary part of a Green's function at finite temperature in the real-time
formalisms. Emphasis is put on the clarification of a recent controversy
concerning these rules in the ``$1/2$" formalism, more precisely on the issue
related to the interpretation of these rules in terms of cut diagrams, like at
$T=0$. On the second hand, new
results are presented, enabling one to calculate the imaginary part of thermal
Green's functions in other formulations of the real-time formalism, like the
``retarded/advanced" formalism in which a lot of simplifications occur.

\end{abstract}
\vskip 4mm
\centerline{\hfill ENSLAPP--A--639/97\hglue 2cm}
\centerline{\hfill hep-ph/9701410\hglue 2cm}
\vfill
\thispagestyle{empty}
\end{titlepage}

\section{Introduction}

A long time ago, Kobes and Semenoff (KS in the following)  presented in a
series of two papers a generalization of the well known Cutkosky's rules in
order to calculate the imaginary part of a Green's function in the Real Time
Formalism (RTF)  at finite temperature. In the first paper \cite{KobesS1}, they
generalized  in a straightforward way the rules valid at zero temperature for
each term  of the sum over the type $1$ or $2$ for the internal vertices of the
diagram. This first step resulted in a huge number of terms to be evaluated in
order to calculate the imaginary part of a diagram at finite temperature. In
their second paper \cite{KobesS2}, they focused on thermal Green's functions
whose external points are all of type $1$, because in these days it was
strongly believed that the type $1$ fields were ``physical", whereas the type
$2$ fields were supposed to be ``ghost fields" that could not appear on the
external lines of a physical amplitude. For this kind of Green's functions, the
rules obtained in their first paper can be simplified considerably, but their
conclusion is that the standard interpretation of these rules in terms of cut
diagrams do not survive the thermalization. Na\"\i vely, the reason of that
difference between the $T=0$ and the $T\not= 0$ cases lies in the fact that at
$T=0$, a cut propagator, proportional to $\theta(\pm k_0)$, imposes a definite
sign to the energy flow, whereas at $T\not=0$, it is proportional to
$\theta(\pm k_0)\pm n_{_{B,F}}(|k_0|)$ so that the energy flow is not
constrained any longer.

Recently, Bedaque, Das and Naik (BDN) claimed \cite{BedaqDN1}  that the
interpretation in terms of cut diagrams holds also at finite temperature,
despite the apparent lack of energy flow constraints in the high temperature
cut propagators. Their proof is essentially based on the Kubo-Martin-Schwinger
(KMS) identities verified by the high temperature Green's functions, which
imply a cancelation of the terms that do not correspond to a cut trough the
diagram when performing the sum over the types $1$ or $2$ of the internal
vertices. It is also obvious that such a cancelation is quite impossible in
KS's rules, since this sum is already performed by these rules. Moreover, a lot
of actual calculations have been carried with the rules of KS, and
contradictions with the results provided by other methods, like calculating the
whole diagram and then taking the imaginary part or calculating the same
quantity in the imaginary time formalism, have never been found, indicating
that KS's rules should be true despite their lack of interpretation in terms
of cut diagrams. Nevertheless, BDN never explained in their paper the reason of
the discrepancy between their result and KS's result.

For these reasons, it seems useful to investigate the precise
connections between the two approaches, in order to explain carefully the
origin of their apparent discrepancy. This is the purpose of the section 
\ref{form12}. The outline of that section is as follows: first, we recall the
common starting point of both methods, which is a direct generalization of the
$T=0$ Cutkosky's rules. Then, we reproduce the derivation of KS's rules,
explaining how terms that does not correspond to cut diagrams arise, and why
they are necessary in this approach. In the paragraph \ref{BDNform}, we expose
an abbreviated version of BDN's proof, leading to a set of cutting rules in
which every term is interpretable as a cut diagram. Finally, we explain where
the apparent contradiction lies and compare the efficiency of both methods on
an explicit example.

The section \ref{other} is devoted to the presentation of the cutting rules
that must be used in other versions of the real-time formalism. Indeed,
Aurenche and Becherrawy \cite{AurenB1}, followed by van Eijck and van Weert
\cite{EijckW1}, first presented a new set of Feynman rules that can be obtained
trough a ``rotation" applied to the usual  ``1/2" formalism. The aim of that
transformation is to obtain a matrix propagator with two vanishing components,
and without statistical factors. The statistical weights are then rejected into
the vertices. The two non-vanishing components are nothing but the retarded and
advanced zero-temperature propagators, which is at the origin of the
denomination of this formalism as ``Retarded/Advanced" (R/A) formalism. Then,
van Eijck, Kobes and  van Weert generalized that kind of transformations in
\cite{EijckKW1}.  The ``R/A" formalism appeared to be only a particular example
of this kind of transformations. In particular, two other formulations of the
RTF appeared to be quite simple: the Keldysh formalism which is useful in
problems related to the linear response theory, and the ``Feynman/Anti Feynman"
($F$/$\bar{F}$) formalism which have also only two non-vanishing propagator
components, which are the usual zero-temperature Feynman propagator and its
complex conjugate.

The ``R/A" formalism then proved to be quite efficient for carrying actual
calculations \cite{AurenBP1}, and to have very close connections with the
imaginary time formalism \cite{Gueri1,Gueri2}. This is the reason why it would
be very useful to have some rules enabling one to calculate the imaginary part
of a Green's function directly in the ``R/A" formalism, and more generally in
all the formalisms of  \cite{EijckKW1}. This will be done quite simply in the
section  \ref{other}, first in the case of general transformations for which
the ``rotation" matrix is real, and then in the particular cases that
correspond to the three formalisms evocated above. The ``R/A" cutting rules
appear to be very simple, and lead to quite compact calculations.  Some
emphasis will be put in the calculation of the imaginary part of self-energy
diagrams, since this quantity is related to the decay rate and production rate
of the external particle. In that particular case, the ``R/A" cutting rules
become even more simple.

Finally, section \ref{concl} is devoted to a summary and to concluding remarks.

\section{Various approaches in the ``$1/2$" formalism}
\label{form12}
\subsection{Common starting point}
We follow here the method which has been derived first by KS in
\cite{KobesS1}. 
In the $1/2$ formalism, every Green's function with external points of types
$a_i=1,2$ can be formally written as the sum over the types $1$ or $2$ of the
internal vertices $v_i$\footnote{In this paper, the momenta of $n$-point Green
functions are defined to be entering. The only exeption to this rule is the
$2$-point functions for which we don't write explicitly the two momenta $k_1$
and $k_2$ since they are related simply by $k_1+k_2=0$. The single momentum
indicated as the argument of $2$-point functions is the one which enters by the
leg corresponding to the first label.}:
\begin{equation}
G^{\{a_i\}}(\{k_i\})=\sum\limits_{\{v_i=1,2\}}^{}G^{\{a_i\}\{v_i\}}(\{k_i\})\; ,
\end{equation}
where $G^{\{a_i\}\{v_i\}}(\{k_i\})$ is the term with external points of fixed
type $a_i$ and internal vertices of type $v_i$. Then, taking the imaginary
part:
\begin{equation}
{\rm Im}\left(iG^{\{a_i\}}(\{k_i\})\right)
=\sum\limits_{\{v_i=1,2\}}^{}
{\rm Im}\left(iG^{\{a_i\}\{v_i\}}(\{k_i\})\right)\; .
\label{sum1}
\end{equation}
The first step towards the rules enabling to perform directly the calculation
of this imaginary part is common to both KS and BDN approaches. It consists
essentially in reproducing what is done at $T=0$ for each term of the sum of
Eq.~(\ref{sum1}) and is based on the largest time equation. Following
\cite{KobesS1}, we are led to introduce two kind of vertices; the usual ones
\begin{equation}
g^1=-ig,\; g^2=+ig\; ,
\end{equation}
and a set of ``circled" vertices (represented by a small circle in the
figures, and with an underlined label in the equations)
\begin{equation}
g^{\underline{1}}=+ig,\; g^{\underline{2}}=-ig\; ,
\end{equation}
which are nothing but the complex conjugates of the standard ones. Of course,
we need also all the propagators enabling one to connect every pair of
vertices. The usual thermal propagator $G^{ab}(k)$ is the propagator connecting
an uncircled vertex of type $a$ to an uncircled vertex of type $b$, the
momentum
flow being $k$ in the direction $a\to b$:
\begin{equation}
G^{ab}(k)=\pmatrix{G^{11}(k)& G^{12}(k)\cr G^{21}(k)&G^{22}(k)\cr}\; ,
\end{equation}
the explicit expressions of these propagators being (for a contour parameter
chosen to be $\sigma=0$) for bosonic fields:
\begin{eqnarray}
&&G^{11}(k)=i{\cal
P}\left({1\over{k^2-m^2}}\right)
+\pi(1+2n_{_{B}}(|k_0|))\delta(k^2-m^2)\\
&&G^{22}(k)=-i{\cal
P}\left({1\over{k^2-m^2}}\right)
+\pi(1+2n_{_{B}}(|k_0|))\delta(k^2-m^2)\\
&&G^{12}(k)=2\pi(\theta(-k_0)+n_{_{B}}(|k_0|))\delta(k^2-m^2)\\
&&G^{21}(k)=2\pi(\theta(k_0)+n_{_{B}}(|k_0|))\delta(k^2-m^2)\; .
\end{eqnarray}
The propagators $G^{\underline{a}b}(k)$ connecting the circled vertex of type
$a$ to the uncircled vertex of type $b$ and carrying the momentum $k$ are
related to the standard ones by:
\begin{equation}
G^{\underline{a}b}(k)=
\pmatrix{G^{12}(k)& G^{12}(k)\cr G^{21}(k)&G^{21}(k)\cr}\; .
\label{circuncirc}
\end{equation}
With now obvious notations, we have also:
\begin{equation}
G^{a\underline{b}}(k)=
\pmatrix{G^{21}(k)& G^{12}(k)\cr G^{21}(k)&G^{12}(k)\cr}\; 
\label{uncirccirc}
\end{equation}
and
\begin{equation}
G^{\underline{a}\underline{b}}(k)=
\pmatrix{G^{22}(k)& G^{12}(k)\cr G^{21}(k)&G^{11}(k)\cr}\; .
\end{equation}
Then, following \cite{KobesS1,BedaqDN1}, we obtain:
\begin{equation}
{\rm Im}\left(iG^{\{a_i\}\{v_i\}}\right)(\{k_i\})=-{1\over
2}\sum\limits_{(\{a_i\}\{v_i\})}^{}G^{(\{a_i\}\{v_i\})}(\{k_i\})\; ,
\end{equation}
where $\sum_{(\{a_i\}\{v_i\})}$ is a sum running over all the possibilities of
circling the points $a_i$ and $v_i$, excepted the two terms where all or none of
the points are circled, and $G^{(\{a_i\}\{v_i\})}(\{k_i\})$ is the
corresponding term. In some sense, this formula is a simple generalization of
the standard $T=0$ Cutkosky's formula. The only difference lies of course in
the fact that at $T=0$, one can further simplify the calculation of the
imaginary part since the energy flow constraints induced by the propagators
connecting circled to uncircled vertices (called ``cut propagators" in the
following) eliminates all the configurations of circles that do not correspond
to a cut trough the diagram. As already mentioned, this appears to be
impossible at high $T$ since the ``cut propagators" do not constrain the energy
flow at non vanishing $T$. Adding now the sum over the types of the internal
vertices, we get for the imaginary part of the complete Green's function:
\begin{equation}
{\rm Im}\left(iG^{\{a_i\}}(\{k_i\})\right)=-{1\over
2}\sum\limits_{\{v_i=1,2\}}^{}\sum\limits_{(\{a_i\}\{v_i\})}^{}
G^{(\{a_i\}\{v_i\})}(\{k_i\})\; .
\label{common}
\end{equation}
This formula is the common starting point of KS and BDN approaches. It is
important to notice that the number of terms to be evaluated in the right hand
side of this equation is quite huge. Indeed, not only the sum over the types
$1$ or $2$ of all the internal vertices is to be performed, but also the sum
over the possibilities of circling vertices, without excluding the circlings
that do not correspond to a cut diagram.

\subsection{Kobes and Semenoff approach}
At that point, KS managed in \cite{KobesS2} to get rid of the sum
$\sum_{\{v_i=1,2\}}$ to obtain a simpler expression of the imaginary part in
the case where all the external points are of type $1$. The reason for such a
choice was based on the common belief according to which the ``physical fields"
were of type $1$, whereas the type $2$ fields were ``ghost fields". Any
physical amplitude therefore should have only type $1$ external points.
Nowadays, this interpretation of type $1$ or $2$ fields has been ruled out by
the fact that it is not clear what a ``physical amplitude" at finite $T$ is
(for instance, the most physically relevant self-energy function is not
$\Sigma^{11}$ but the function which appears in the pole shift, as can be seen
by performing a Dyson summation). Nevertheless, in the case of self-energies,
the imaginary part of the relevant self-energy is related to the imaginary part
of $\Sigma^{11}$ in a simple way (see \cite{KobesS2}), so that the calculation
of ${\rm Im}(\Sigma^{11})$ is sufficient.

In their final formula, there are only type $1$ external points and internal
vertices:
\begin{equation}
{\rm Im}\left(iG^{\{a_i=1\}}(\{k_i\})\right)=- {1\over
2}\sum\limits_{{(\{a_i=1\})}\atop{[\{v_i=1\}]}}^{}
G^{(\{a_i=1\}),[\{v_i=1\}]}(\{k_i\})\;
,
\label{KSfor}
\end{equation}
where the sum $\sum_{{(\{a_i=1\})}\atop{[\{v_i=1\}]}}$ runs over all the
possibilities of circling the external points $a_i$ excepted the two terms where
all or none of them are circled, and {\it all} the possibilities of circling
the internal vertices $v_i$. The important advantage compared to
Eq.~(\ref{common}) lies in the fact that the sum over the type of the internal
vertices is already performed, which reduces drastically the number of terms to
be evaluated. A very simple proof of that formula can also be found in
\cite{Althe1}.

This formula leads to some difficulties in its interpretation since it
generates some terms that does not correspond to cuts through the diagram, {\it
i.e.} to topologies where the ``cut propagators" (propagators connecting a
circled to an uncircled vertex) do not divide the diagram in two connected
parts. two examples of such a contributions are reproduced in Fig.~\ref{uncut}.
One should note that in this approach, there is a complete equivalence between
the cut propagators and the ``on-shell" propagators, {\it i.e.} propagators
that are proportional to a factor $\delta(k^2-m^2)$. Moreover, it should be
emphazised that these ``uncut terms" are necessary in this formula in order to
have the cancelation of the pathologies consisting in products of distributions
having overlapping supports, like $\delta^2(k^2-m^2)$ or $({\cal
P}(1/(k^2-m^2)))^2$: without these terms, the result would be ill-defined. This
is what happens for the first example of Fig.~\ref{uncut}. But, it would not be
consistent to simply drop these ``uncut terms", by arguing that they are here
only for the cancelation of pathologies, because that these ``uncut terms" can
occur even when there are no pathologies to be canceled, like in the second
diagram of Fig.~\ref{uncut}.

So, the conclusion of this approach is that despite a considerable reduction of
the number of terms, the formula Eq.~(\ref{KSfor}) generates unavoidably a few
terms that are not interpretable in terms of cut diagrams, these terms being
absolutely necessary for the completeness of the result.

\subsection{Bedaque, Das and Naik approach}
\label{BDNform}
After formula (\ref{common}), BDN took an orthogonal way compared to KS, since
they chose to reduce the number of terms contained in that formula by trying
to simplify the sum $\sum_{(\{a_i\}\{v_i\})}$ while keeping the sum
$\sum_{\{v_i=1,2\}}$ unchanged. In their approach, there is no need of a
special assumption concerning the type $1$ or $2$ of the external points of the
diagram. In fact, they proved that the effect of the sum $\sum_{\{v_i=1,2\}}$
was to kill all the configurations of circlings that do not correspond to cut
diagrams, a cut diagram being generically of the type depicted in
Fig.~\ref{cut}. Of course, in the remaining terms, the sum $\sum_{\{v_i=1,2\}}$
is still to be performed.

In what follows, we give a simple proof of that result. Let us first notice
that it is enough to prove that the configurations where a set of uncircled
vertices is completely surrounded by circled vertices only (and vice versa) are
killed by the sum over the type of the $v_i$'s. These configurations occur when
the disposition of the circles in the diagram is such that the diagram contains
a subdiagram of one of the types depicted in Fig.~\ref{cancelled}. In the
figure~\ref{cancelled}, the vertices labeled by $c_i$ are obviously internal,
so that we must sum over the types $c_i=1,2$. It is this sum that will make
these configurations vanish.

Let us begin by the configuration of Fig.~\ref{cancelled}-(a). The
corresponding terms in the right hand side of Eq.~(\ref{common}) are
proportional to the quantity:
\begin{equation}
F^{\{\underline{b_i}\}}(\{k_i\})\equiv \sum\limits_{\{c_i=1,2\}}^{}
\Gamma^{\{c_i\}}(\{k_i\})\, \prod\limits_{i=1}^{n}
G^{\underline{b_i}c_i}(k_i)\; ,
\end{equation}
where $\Gamma^{\{c_i\}}(\{k_i\})$ is the $n$-point vertex function with
external points $c_i$ and entering momenta $k_i$ (the signs associated to type
$c_i=2$ external vertices are included in its definition).
Looking at the propagators connecting a circled vertex to an uncircled one in 
Eq.~(\ref{circuncirc}), we obtain:
\begin{equation}
F^{\{\underline{b_i}\}}(\{k_i\})=\prod\limits_{\{i|b_i=1\}} G^{12}(k_i)
\prod\limits_{\{i|b_i=2\}} G^{21}(k_i) \sum\limits_{\{c_i=1,2\}}
\Gamma^{\{c_i\}}(\{k_i\}) =0 \; ,
\end{equation}
since the $n$-point vertex functions satisfy the identity
\cite{ChouSHY1,Evans5}:
\begin{equation}
\sum\limits_{\{c_i=1,2\}}
\Gamma^{\{c_i\}}(\{k_i\}) =0\; .
\label{constraint1}
\end{equation}
This identity is in fact the Fourier expression of the largest time equation
used in \cite{KobesS1}.
It is worth noticing that this cancelation survives in an
out-of-equilibrium\footnote{Here and in the following, when we speak of an
``out-of-equilibrium field theory", we have simply in mind the same theory
where the Kubo-Martin-Schwinger relations are not satisfied. This is the case
in the simple modification of the real time formalism that consists in
replacing the Bose-Einstein statistical factor by an arbitrary function. Of
course, we are aware of the immediate difficulties one is faced to when
applying such a formalism \cite{AltheS1,Althe6,Bedaq1,BellaM2}, so that our
remarks about non equilibrium theories should be understood only as indications
of what a theory without the KMS identity would look like and not as a claim
of what should be a theory of non equilibrium systems.} field theory since this
identity is always verified.

Considering now the configuration of Fig.~\ref{cancelled}-(b), the
corresponding terms in the right hand side of Eq.~(\ref{common}) are
proportional to:
\begin{equation}
F^{\{b_i\}}(\{k_i\})\equiv \sum\limits_{\{c_i=1,2\}}^{}
\Gamma^{\{\underline{c_i}\}}_{\rm circled}(\{k_i\})\, \prod\limits_{i=1}^{n}
G^{b_i\underline{c_i}}(k_i)\; ,
\end{equation}
where $\Gamma^{\{\underline{c_i}\}}_{\rm circled}(\{k_i\})$ is the $n$-point
vertex function made only of circled vertices. Therefore, this vertex function
is nothing but the complex conjugate of the previous one:
\begin{equation}
\Gamma^{\{\underline{c_i}\}}_{\rm circled}(\{k_i\})=
\left(\Gamma^{\{c_i\}}(\{k_i\})\right)^{*}\; .
\end{equation}
Making use of Eq.~(\ref{uncirccirc}), we can write:
\begin{Eqnarray}
F^{\{b_i\}}(\{k_i\})&&=\sum\limits_{\{c_i=1,2\}} \left(
\Gamma^{\{c_i\}}(\{k_i\})\right)^{*} \prod\limits_{\{i|c_i=1\}} G^{21}(k_i)
\prod\limits_{\{i|c_i=2\}}  G^{12}(k_i)\nonumber\\
&&=\left[\prod\limits_{i=1}^{n} 2\pi(\theta(k_i^0)+n_{_{B}}(|k_i^0|))
\delta(k_i^2-m_i^2)\right]
\;\sum\limits_{\{c_i=1,2\}} \left(\prod\limits_{\{i|c_i=2\}}
e^{-\beta k_i^0}\right) \left(\Gamma^{\{c_i\}}(\{k_i\})\right)^{*}\nonumber\\
&& =0\; ,
\end{Eqnarray}
the last equality being a consequence of the identity \cite{ChouSHY1,Evans5}:
\begin{equation}
\sum\limits_{\{c_i=1,2\}} \left(\prod\limits_{\{i|c_i=2\}}
e^{-\beta k_i^0}\right) \Gamma^{\{c_i\}}(\{k_i\})=0\; ,
\label{KMS}
\end{equation}
which is nothing but the expression of the Kubo-Martin-Schwinger property for
$n$-point real time vertex functions. This relation was used by \cite{KobesS2} 
in time coordinates where it appears to be related to the $-i\beta$ periodicity
properties of Green's functions. One can note that this cancelation is not true
in an out-of-equilibrium plasma since it makes use of the KMS relations which
are specific to thermal equilibrium.

Therefore, taking these cancelations into account, we can reduce the sum 
$\sum_{(\{a_i\}\{v_i\})}$ over the disposition of circles to a smaller sum
where we keep only those terms that correspond to a cut through the diagram,
like on Fig.~\ref{cut}. We can then write:
\begin{equation}
{\rm Im}\left(iG^{\{a_i\}}(\{k_i\})\right)=- {1\over 2}
\sum\limits_{\{v_i=1,2\}} \sum\limits_{\rm cuts} 
G^{(\{a_i\}\{v_i\})}(\{k_i\})\; .
\label{BDNfor}
\end{equation}
In that formula, the sum over the type $1$ or $2$ of the internal vertices
$v_i$ is still to be performed for the remaining terms. This fact happens to
generate more terms than with the final formula of Kobes and Semenoff. The
issue of the efficiency of both methods will be discussed on an example in the
next paragraph.

One could then wonder why it is possible thanks to the KMS relations to
eliminate the same configurations of circles than at $T=0$, despite the the
fact that apparently these relations have nothing to do with energy flow
constraints. In fact, the study of the $T=0$ limit of the KMS relations of
Eq.~(\ref{KMS}) indicates that the KMS identity appears to be a thermal
generalization of some energy flow constraints. Indeed, it is possible to show
that a direct consequence of Eq.~(\ref{KMS}) is:
\begin{equation}
\lim_{T\to 0^{+}} \Gamma^{\{c_i\}}(\{k_i\})\propto
\theta\left(\sum\limits_{\{i|c_i=2\}} k_i^0\right)\; .
\end{equation}
A proof of this relation is given in the appendix. So, even if this limit is
not a proof of the fact that KMS should play at positive $T$ the same role than
energy flow constraints at $T=0$, it points towards the interpretation of the
KMS identity as the thermal generalization of a very simple  constraint on
energy flow\footnote{As a by-product, this $T\to 0^+$ limit leads to a trivial
proof of the fact that the real time ``1/2" formalism goes into the standard
$T=0$ Feynman rules ({\it i.e.} without the need of a matrix propagator) when
one is calculating a function with only type $1$ external points. Indeed, when
inserting a vertex function between $n$ legs, we have
\begin{equation}
\lim_{T\to 0^+} \sum_{\{c_i=1,2\}}\Gamma^{\{c_i\}}(\{k_i\})
\prod\limits_{i=1}^{n} G^{1c_i}(k_i) = \Gamma^{\{1\cdots 1\}}_{T=0}(\{k_i\})
\prod\limits_{i=1}^{n} G^{11}_{T=0}(k_i)\; ,
\end{equation}
and we prove by induction that only type $1$ vertices are needed.}.

\subsection{Are the two methods contradictory ?}
Let us first explain where lies the apparent contradiction between the
conclusions of KS and BDN concerning the issue of the interpretation in
terms of cut diagrams:
\begin{itemize}
\item[KS:] In Eq.~(\ref{KSfor}), some terms are not interpretable
as the contribution of a cut diagram. Moreover, as already said, the uncut
propagators can only be $G^{11}(k)$ or
$G^{\underline{1}\underline{1}}(k)=G^{22}(k)$. It means that the uncut
propagators are always off-shell ({\it i.e.} not proportional to a Dirac delta
function). On the contrary, all the cut propagators are on-shell.

\item[BDN:] In Eq.~(\ref{BDNfor}), all the terms are
interpretable as contributions of cut diagrams. Moreover, with the same
definition as in KS approach for a cut propagator ({\it i.e.} connecting the
circled vertices to the uncircled ones), the uncut propagators in this approach
can be any of $G^{ab}(k)$ or $G^{\underline{a}\underline{b}}(k)$. Among them,
we find $G^{12}(k)$ and $G^{21}(k)$ which are on-shell. This is an essential
difference with the approach of KS.
\end{itemize}
Therefore, the expression ``cut propagator" does not cover the same reality in
both approaches. Some of the uncut propagators (more precisely, those which are
on-shell) of BDN are considered to be cut ones in KS method. It means that more
propagators are called ``cut propagators" in KS's approach than in BDN's one,
and this is of course at the origin of the apparent discrepancy. Na\"\i vely
speaking, if one has more ``cut propagators", then it will be difficult
 to have only configurations where the set of cut propagators form a
single cut through the diagram. This assertion is illustrated in the figure
\ref{hidden}, in which we see clearly how some of the cut propagators of KS,
spoiling the interpretation of the imaginary part in terms of cut diagrams, are
``hidden" in the approach of BDN.

Now, in order to be more explicit, let us consider in detail the example of the
diagram of Fig.~\ref{uncut} for the theory of a real scalar field.
 If we use the method of KS to compute the
contribution of this topology to the imaginary part of $\Sigma^{11}(k)$, we
must evaluate the $8$ terms represented on Fig.~\ref{KS}. After some
straightforward\footnote{To obtain
 these expressions, we get rid of two components
of the one-loop self-energy $\sigma^{ij}(k)$ 
by using the two identities they verify:
\begin{eqnarray}
&&\sigma^{11}(k)+\sigma^{22}(k)+\sigma^{12}(k)+\sigma^{21}(k)=0\\
&&\sigma^{11}(k)+\sigma^{22}(k)+e^{\beta k^0}\sigma^{12}(k)+
e^{-\beta k^0}\sigma^{21}(k)=0\; .
\end{eqnarray}
Moreover, we use the following relations
\begin{eqnarray}
&&{\cal P}{1\over{x^2}}=\left({\cal P}{1\over x}\right)^2-\pi^2\delta^2(x)\\
&&2\delta(x){\cal P}{1\over x}=-\delta'(x)\; ,
\end{eqnarray}
relating the various distributions appearing in this calculation.} 
calculations, we get for the contribution of the
individual terms (actually, we grouped them by pairs since some terms
differ only by a factor $e^{-\beta q^0}$, as proven in \cite{KobesS2}):
\begin{eqnarray}
{\rm Im}^{^{KS}}_{(1+5)}(\Sigma^{11}(q))=&&
{{(1+e^{-\beta q^0})}\over{2}}
\int{{d^4r}\over{(2\pi)^4}}\;
G^{21}(p)\left[{{(1+e^{\beta r^0})\sigma^{21}(r)+e^{\beta r^0}\sigma^{22}(r)}
\over{e^{\beta r^0}-1}}\right]\nonumber\\
&&\times\left\{2i\pi^2{{e^{\beta r^0}+1}
\over{e^{\beta r^0}-1}}\delta^2(r^2-m^2)
+\pi\epsilon(r^0)\delta'(r^2-m^2)\right\}\; ,\\
{\rm Im}^{^{KS}}_{(2+6)}(\Sigma^{11}(q))=&&
-{{(1+e^{-\beta q^0})}\over{2}}
\int{{d^4r}\over{(2\pi)^4}}\;
G^{21}(p)\sigma^{21}(r)\nonumber\\
&&\times\left\{2i\pi^2{{e^{2\beta r^0}+1}
\over{(e^{\beta r^0}-1)^2}}\delta^2(r^2-m^2)
+i{\cal P}{1\over{(r^2-m^2)^2}}\right\}\; ,\\
{\rm Im}^{^{KS}}_{(3+7)}(\Sigma^{11}(q))=&&
{{(1+e^{-\beta q^0})}\over{2}}
\int{{d^4r}\over{(2\pi)^4}}\;
{{e^{\beta r^0}G^{21}(p)\sigma^{22}(r)}\over{e^{\beta r^0}-1}}\nonumber\\
&&\times\left\{-2i\pi^2{{e^{\beta r^0}+1}
\over{e^{\beta r^0}-1}}\delta^2(r^2-m^2)
+\pi\epsilon(r^0)\delta'(r^2-m^2)\right\}\; ,\\
{\rm and}\nonumber\\
{\rm Im}^{^{KS}}_{(4+8)}(\Sigma^{11}(q))=&&
-{{(1+e^{-\beta q^0})}\over{2}}
\int{{d^4r}\over{(2\pi)^4}}\;
{{e^{\beta r^0}G^{21}(p)\sigma^{21}(r)}\over{(e^{\beta
r^0}-1)^2}}4i\pi^2\delta^2(r^2-m^2)\; ,
\end{eqnarray}
where the $\sigma^{ij}(r)$ are components of the one-loop self-energy matrix.
It is instructive to note that the cancelation of the pathologic products of
distributions occur only when all the terms are added, and not at the level
of the individual terms.
Adding all these contributions, we finally get\footnote{Thanks to the spectral
representation of 2-point vertex functions (see \ref{self}), it is easy to
prove that $\sigma^{21}(r)$ is imaginary and that the quantity $2e^{\beta
r^0}\sigma^{22}(r)+(1+e^{\beta r^0})\sigma^{21}(r)$ is real, so that the sum of
all the contributions is obviously real.}:
\begin{Eqnarray}
{\rm Im}^{^{KS}}_{(1+\cdots+8)}(\Sigma^{11}(q))=&&
{{(1+e^{-\beta q^0})}\over{2}}
\int{{d^4r}\over{(2\pi)^4}}\;
G^{21}(p)\left\{-i\sigma^{21}(r){\cal P}{1\over{(r^2-m^2)^2}}\right.\nonumber\\
&&\left.+\pi\left[{{2e^{\beta r^0}\sigma^{22}(r)+(1+e^{\beta
r^0})\sigma^{21}(r)}\over{e^{\beta r^0}-1}}
\right]\epsilon(r^0)\delta'(r^2-m^2)\right\}
\; .
\label{exKS}
\end{Eqnarray}

For the same quantity, the method of BDN requires the calculation of the
$6\times4=24$
terms of Fig.~\ref{BDN}. For each disposition of the circles, the
calculations are heavier than in the KS's method since one should perform the
sum over $a,b=1,2$. Here again, we can group the terms by pairs:
\begin{eqnarray}
{\rm Im}^{^{BDN}}_{(1+6)}(\Sigma^{11}(q))=&&
-{{(1+e^{-\beta q^0})}\over{2}}
\int{{d^4r}\over{(2\pi)^4}}\;
G^{21}(p)\left[{{e^{\beta r^0}\sigma^{22}(r)+\sigma^{21}(r)}\over{e^{\beta
r^0}-1}}\right]\nonumber\\
&&\times
\left\{2i\pi^2\delta^2(r^2-m^2)-\pi\epsilon(r^0)
\delta'(r^2-m^2)\right\}\; ,\\
{\rm Im}^{^{BDN}}_{(2+5)}(\Sigma^{11}(q))=&&
-{{(1+e^{-\beta q^0})}\over{2}}
\int{{d^4r}\over{(2\pi)^4}}\;
G^{21}(p)\sigma^{21}(p)\nonumber\\
&&\times
\left\{2i\pi^2\delta^2(r^2-m^2)+i{\cal P}
{1\over{(r^2-m^2)^2}}\right\}\; ,\\
{\rm and}\nonumber\\
{\rm Im}^{^{BDN}}_{(3+4)}(\Sigma^{11}(q))=&&
{{(1+e^{-\beta q^0})}\over{2}}
\int{{d^4r}\over{(2\pi)^4}}\;
G^{21}(p)\left[{{e^{\beta r^0}(\sigma^{22}(r)+\sigma^{21}(r))}\over{e^{\beta
r^0}-1}}\right]\nonumber\\
&&\times
\left\{2i\pi^2\delta^2(r^2-m^2)+\pi\epsilon(r^0)\delta'(r^2-m^2)\right\}\; .
\end{eqnarray}
Of course, the sum of all the contributions is the same as the one obtained
previously in Eq~(\ref{exKS}). Here again, we note that if we sum only groups
of terms that correspond to the same cut, we obtain expressions that are not
defined by themselves. This seems to indicate that, even if in the approach of
Bedaque, Das and Naik the interpretation in terms of cut diagrams survives, the
interpretation of the terms obtained as products of physical amplitudes is
meaningless since these terms are ill-defined when considered alone. This
conclusion is different from that of BDN, but this discrepancy is simply
related to the fact that they never considered in \cite{BedaqDN1} explicit
examples where pathologies do arise.

To conclude this section, one can say that the methods provided by KS and BDN
are not in contradiction despite the appearances; that KS's method leads to
simpler calculations; and that, contrary to the claim of BDN, the decomposition
provided by Eq.~(\ref{BDNfor}) is not completely consistent with an
interpretation in terms of the underlying physical amplitudes.

\section{About the other bases of the real-time formalism}
\label{other}
\subsection{General considerations}
\label{gener}
We follow here the approach of \cite{EijckKW1} in order to define new
formulations of the real time formalism, initially expressed in the $1/2$
base. The general aim of this method is to derive formalisms in which the
constraints provided by the relations of Eq.~(\ref{constraint1}) and
Eq.~(\ref{KMS}) correspond to the nullity of certain components of the new
$n$-point vertex functions, instead of the nullity of complicated linear
combinations. Another practical advantage lies in the fact that for some of
these modified formalisms the statistical weights are rejected into the
vertices, which clarify the calculations.

Let us define new Green's functions by the transformation:
\begin{equation}
{\imb G}^{\{X_i\}}(\{k_i\})\equiv\sum\limits_{\{a_i=1,2\}}
G^{\{a_i\}}(\{k_i\})\prod\limits_{i=1}^{n}U^{X_i a_i}(k_i) \; ,
\label{defrot}
\end{equation}
with an invertible ``rotation matrix" $U^{Xa}(k)$ so that the reverse
transformation exists. The new index $X$ takes also two distinct values, and
will be denoted by capital letters. Before going further, it is useful to
slightly modify the notations for the vertices in the $1/2$ formalism. Instead
of $g^1$ and $g^2$, we introduce $8$ vertices $g^{abc}$, so that
$g^{111}=-g^{222}=-ig$ whereas all the other components are vanishing. By doing
so, at each vertex, we have to perform the sum $\sum_{\{a,b,c=1,2\}}g^{abc}$.
In the new formalism, the vertex functions should be defined so that holds the
relation:
\begin{equation}
\sum\limits_{\{X_i\}}^{}{\imb \Gamma}^{\{X_i\}}(\{k_i\}) 
\prod\limits_{i=1}^{n}
{\imb G}^{Y_iX_i}(k_i)=\sum\limits_{\{a_i=1,2\}}\sum\limits_{\{b_i=1,2\}}
\Gamma^{\{b_i\}}(\{k_i\}) \prod\limits_{i=1}^{n} U^{Y_ia_i}(k_i)
G^{a_ib_i}(k_i)\; ,
\end{equation}
in order to be consistent with the definition chosen for complete 
Green's functions. By multiplication of the above equation by the appropriate
inverse propagators, this requirement leads one to:
\begin{equation}
{\imb \Gamma}^{\{X_i\}}(\{k_i\})=\sum\limits_{\{a_i=1,2\}}
\Gamma^{\{a_i\}}(\{k_i\}) \prod\limits_{i=1}^{n} V^{X_ia_i}(k_i)\; ,
\label{rotvertex}
\end{equation}
where we defined
\begin{equation}
V^{Xa}(k)\equiv \left((U^{^{T}})^{-1}\right)^{Xa}(-k)\; .
\end{equation}

Now, in order to be able to say something about the imaginary part of the
transformed Green's functions, we limit ourselves to those transformations for
which the rotation matrix $U(k)$ is real. With such a restriction, we have:
\begin{equation}
{\rm Im}\left(i{\imb G}^{\{X_i\}}(\{k_i\})\right)=\sum\limits_{\{a_i=1,2\}}
{\rm Im}\left(iG^{\{a_i\}}(\{k_i\})\right) \prod\limits_{i=1}^{n}
U^{X_ia_i}(k_i)\; .
\end{equation}

Then, we reproduce the usual way of justifying the Feynman's rules in the
transformed formalism, which consists in writing each object (propagator,
vertex) belonging to the $1/2$ formalism in the right hand side of
Eq.~(\ref{defrot}) as a function of its counterpart in the rotated formalism.
Then, the rotation matrices cancel by pairs $UU^{-1}$ when linking propagators
to vertices. If we apply this procedure to the circling rules obtained in the
``$1/2$" formalism for the calculation of ${\rm Im}\left(G^{\{a_i\}}\right)$,
we obtain analogous circling rules in the transformed formalism. For this, it
is necessary to define, in complete analogy with the definition of the
transformed propagators and vertices, the counterparts of $G^{\underline{a}b}$,
$G^{a\underline{b}}$, $G^{\underline{a}\underline{b}}$ and
$g^{\underline{a}\underline{b}\underline{c}}$. This gives:
\begin{eqnarray}
&&{\imb G}^{\underline{X}Y}(k)=\sum\limits_{\{a,b=1,2\}} U^{Xa}(k) U^{Yb}(-k)
G^{\underline{a}b}(k)\\
&&{\imb G}^{X\underline{Y}}(k)=\sum\limits_{\{a,b=1,2\}} U^{Xa}(k) U^{Yb}(-k)
G^{a\underline{b}}(k)\\
&&{\imb G}^{\underline{X}\underline{Y}}(k)=\sum\limits_{\{a,b=1,2\}} 
U^{Xa}(k) U^{Yb}(-k)
G^{\underline{a}\underline{b}}(k)
\end{eqnarray}
for the propagators, and:
\begin{Eqnarray}
{\imb g}^{\underline{XYZ}}(k_1,k_2,k_3)&&=\sum\limits_{\{a,b,c=1,2\}}
V^{Xa}(k_1) V^{Yb}(k_2) V^{Zc}(k_3) g^{\underline{abc}}\nonumber\\
&&=-{\imb g}^{XYZ}(k_1,k_2,k_3)
\end{Eqnarray}
for the vertex. 
If we denote by $\alpha$ and $\bar{\alpha}$ the two values taken by the 
index $X$ in the new
formalism, the analogue of Eq.~(\ref{common}) is:
\begin{equation} {\rm Im}\left({\imb G}^{\{X_i\}}(\{k_i\})\right)=-{1\over 2}
\sum\limits_{\{V_i=\alpha\alpha\alpha,\alpha\alpha\bar{\alpha},\cdots\}}
\;\sum\limits_{(\{X_i\}\{V_i\})} {\imb G}^{(\{X_i\}\{V_i\})}(\{k_i\})\; ,
\end{equation}
where the sum
$\sum_{\{V_i=\alpha\alpha\alpha,\alpha\alpha,\bar{\alpha},\cdots\}}$ is the sum
over the possible types of the internal vertices $V_i$, and the sum 
$\sum_{(\{X_i\}\{V_i\})}$ runs over all the possibilities of circling external
points and vertices, excepted the two terms where all or none of them are
circled.

Here also, it is possible to eliminate from the sum all the terms that do not
correspond to a cut diagram. To do so, we need first to obtain the
relations that correspond in the new formalism to the constraints provided by
Eq.~(\ref{constraint1}) and Eq.~(\ref{KMS}). Let us first invert the relation
Eq.~(\ref{rotvertex}), which gives:
\begin{equation}
\Gamma^{\{a_i\}}(\{k_i\})=\sum\limits_{\{X_i=\alpha,\bar{\alpha}\}}
{\imb \Gamma}^{\{X_i\}}(\{k_i\})
\prod\limits_{i=1}^{n}\left(U^{^{T}}\right)^{a_iX_i}(-k_i)\; .
\end{equation}
Then, it is trivial to obtain the relation that corresponds to
Eq.~(\ref{constraint1}) in the new formalism
\begin{equation}
\sum\limits_{\{X_i=\alpha,\bar{\alpha}\}}\left(
\sum\limits_{\{a_i=1,2\}}\prod\limits_{i=1}^{n}
\left(U^{^{T}}\right)^{a_iX_i}(-k_i)\right){\imb \Gamma}^{\{X_i\}}(\{k_i\})=0\; ,
\label{rotconstraint1}
\end{equation}
and for the KMS relation Eq.~(\ref{KMS}):
\begin{equation}
\sum\limits_{\{X_i=\alpha,\bar{\alpha}\}}\left(
\sum\limits_{\{a_i=1,2\}}\prod\limits_{\{i|a_i=2\}} e^{-\beta k_i^0}
\prod\limits_{i=1}^{n}
\left(U^{^{T}}\right)^{a_iX_i}(-k_i)\right){\imb \Gamma}^{\{X_i\}}(\{k_i\})=0\; .
\label{rotKMS}
\end{equation}
To see if the configurations analogous to that of Fig.~\ref{cancelled}-(a) do
cancel in the sum, we are lead to the evaluation of 
\begin{Eqnarray}
{\imb F}^{\{\underline{X_i}\}}(\{k_i\})&&\equiv
\sum\limits_{\{Y_i=\alpha,\bar{\alpha}\}} {\imb
\Gamma}^{\{Y_i\}}(\{k_i\}) \prod\limits_{i=1}^{n} {\imb
G}^{\underline{X_i}Y_i}(k_i)\nonumber\\
&&=\sum\limits_{\{a_i=1,2\}}\prod\limits_{\{i|a_i=1\}} G^{12}(k_i)
\prod\limits_{\{i|a_i=2\}} G^{21}(k_i) \prod\limits_{i=1}^{n} U^{X_ia_i}(k_i)
\nonumber\\
&&\quad\times
\sum\limits_{\{Y_i=\alpha,\bar{\alpha}\}}\left(\sum\limits_{\{b_i=1,2\}}\prod\limits_{i=1}^{n}
\left(U^{^{T}}\right)^{b_iY_i}(-k_i)\right) {\imb
\Gamma}^{\{Y_i\}}(\{k_i\})\nonumber\\
&&=0\; ,
\end{Eqnarray}
thanks to Eq.~(\ref{rotconstraint1}). This cancelation survives in
out-of-equilibrium theories since it is based only on the generalization to the
new formalism of an identity which is always true.
For the configuration of circles of Fig.~\ref{cancelled}-(b), we must now
evaluate:
\begin{Eqnarray}
{\imb F}^{\{X_i\}}(\{k_i\})&&\equiv \sum\limits_{\{Y_i=\alpha,\bar{\alpha}\}} {\imb
\Gamma}^{\{\underline{Y_i}\}}_{\rm circled}(\{k_i\}) \prod\limits_{i=1}^{n}
{\imb G}^{X_i\underline{Y_i}}(k_i)\nonumber\\
&&=\sum\limits_{\{a_i=1,2\}} \prod\limits_{i=1}^{n} 2\pi
(\theta(k_i^0)+n_{_{B}}(|k_i^0|))\delta(k_i^2-m_i^2)U^{X_ia_i}(k_i)
\nonumber\\
&&\quad\times
\sum\limits_{\{Y_i=\alpha,\bar{\alpha}\}}\left(\sum\limits_{\{b_i=1,2\}}
\prod\limits_{\{i|b_i=2\}} e^{-\beta k_i^0} \prod\limits_{i=1}^{n}
\left(U^{^{T}}\right)^{b_iY_i}(-k_i)\right) \left({\imb
\Gamma}^{\{Y_i\}}(\{k_i\})\right)^{*}\nonumber\\
&&=0\; ,
\end{Eqnarray}
due to Eq.~(\ref{rotKMS}). Therefore, the same configurations of circles can be
eliminated from the sum giving the imaginary part of the Green's function in
any formalism obtained by Eq.~(\ref{defrot}). As a consequence, we can
generalize in any such formalism the formula of BDN:
\begin{equation}
{\rm Im}\left({\imb G}^{\{X_i\}}\right)(\{k_i\}) = -{1\over 2}
\sum\limits_{\{V_i=\alpha\alpha\alpha,\alpha\alpha\bar{\alpha},\cdots\}} 
\sum\limits_{\rm cuts} {\imb
G}^{(\{X_i\}\{V_i\})}(\{k_i\})\; .
\label{rotBDN}
\end{equation}

In the following paragraphs, we specialize to three particular such formalisms,
that are the Keldysh formalism, the ``$F/\bar{F}$" formalism and the ``R/A"
formalism.

\subsection{The Keldysh basis}
\label{keld}

The aim of the transformation leading to this formalism is to use the relation
Eq.~(\ref{constraint1}) in order to make one component of the matrix propagator
vanish. Since this identity is true even in out-of-equilibrium theories, this
formalism can be applied without modifications in such a case\footnote{In this
section again, the expression ``out-of-equilibrium" should be interpreted with
great care, since the most straightforward generalization to non-equilibrium
distribution functions of the various
real-time formalisms used here runs immediately into difficulties related to
the non-cancelation of pathologic terms (this is unavoidable since this
cancelation works thanks to the KMS identity). Therefore, when we compare the
equilibrium and ``non-equilibrium" formalisms, we are only studying how
important are the complications when the KMS relation is not satisfied.}.

A ``rotation matrix" leading to that is\footnote{In fact, this rotation matrix
is obtained as a solution of the equation requiring that one
component of the new propagator is proportional to that combination of the
$G^{ij}(k)$ which is always vanishing (the analogue of the left hand side of
Eq.~(\ref{constraint1}) in the case of the propagator).}:
\begin{equation}
U(k)={1\over{\sqrt{2}b(-k^0)}} \pmatrix{&b(k^0)b(-k^0) &b(k^0)b(-k^0)\cr
&+1 &-1\cr}\; ,
\end{equation}
where $b(k^0)$ is an arbitrary function of $k^0$, non vanishing at $k^0=0$.
In this paragraph, we denote by $+$ the label of the first row , 
and by $-$ the label of the second one. 
Performing such a transformation, we get for the propagators:
\begin{eqnarray}
&&{\imb G}^{XY}(k)=\pmatrix{&b(k^0)b(-k^0)(G^{12}(k)+G^{21}(k)) &
G^{11}(k)-G^{12}(k)\cr &G^{11}(k)-G^{21}(k) & 0 \cr}\; ,\\
&&{\imb G}^{\underline{X}Y}(k)=\pmatrix{&b(k^0)b(-k^0)(G^{12}(k)+G^{21}(k)) &
0\cr &G^{12}(k)-G^{21}(k) & 0 \cr}\; ,\\
&&{\imb G}^{X\underline{Y}}(k)=\pmatrix{&b(k^0)b(-k^0)(G^{12}(k)+G^{21}(k)) &
G^{21}(k)-G^{12}(k)\cr & 0 & 0 \cr}\; ,\\
&&{\rm and}\nonumber\\
&&{\imb G}^{\underline{XY}}(k)=\pmatrix{&b(k^0)b(-k^0)(G^{12}(k)+G^{21}(k)) &
G^{21}(k)-G^{11}(k)\cr &G^{12}(k)-G^{11}(k) & 0 \cr}\; .
\end{eqnarray}
As announced, the standard propagator ${\imb G}^{XY}(k)$ possesses a vanishing
component. We see also that the propagators connecting circled to uncircled
vertices have two vanishing components. Despite these zeros, the
simplifications one could expect in the calculations based on this formalism
are almost compensated by the fact that the vertices are more intricate than in
the $``1/2"$ formalism. Indeed, one obtains for the vertices:
\begin{equation}
{{{\imb g}^{X_1X_2X_3}(k_1,k_2,k_3)}\over{-ig}}
={{1-(-1)^{\#\{i|X_i=-\}}}\over{2^{3/2}}}
\prod\limits_{\{i|X_i=+\}}{1\over{b(-k_i^0)}}
\prod\limits_{\{i|X_i=-\}}b(k_i^0)\; ,
\end{equation}
where $\#\{i|X_i=-\}$ stands for the number of indices equal to $-$. It appears
that the function $b(k^0)$, which remained unspecified until now, can simply
be chosen to equal $b(k^0)=1$.

On can note that the extra-diagonal terms of the matrix propagator in the
Keldysh formalism are nothing but retarded and advanced zero-temperature
propagators:
\begin{eqnarray}
&&G^{11}(k)-G^{12}(k)=\Delta_{_{R}}(k)\equiv i{\cal
P}{1\over{k^2-m^2}}+\pi\epsilon(k_0)\delta(k^2-m^2)\\
&&G^{11}(k)-G^{21}(k)=\Delta_{_{A}}(k)\equiv i{\cal
P}{1\over{k^2-m^2}}-\pi\epsilon(k_0)\delta(k^2-m^2)\; .
\end{eqnarray}
For $n$-point functions, 
${\imb G}^{+-\cdots-}(\{k_i\})$ is a $n$-point
retarded function.
More generally, the Green's functions in the Keldysh formalism can be
interpreted as response functions in the linear response theory.

At that point, the formula we get for the calculation of the imaginary part of
a Green's function expressed in the Keldysh basis is, in analogy with
Eq.~(\ref{common}):
\begin{equation}
{\rm Im}\left({\imb G}^{\{X_i\}}(\{k_i\})\right)=-{1\over 2}
\sum\limits_{\{V_i=+++,++-,\cdots\}} \;\sum\limits_{(\{X_i\}\{V_i\})} {\imb
G}^{(\{X_i\}\{V_i\})}(\{k_i\})\; ,
\end{equation}
where $\sum_{\{V_i=+++,++-,\cdots\}}$ is the sum over the possible
types of the internal vertices $V_i$, and the sum  $\sum_{(\{X_i\}\{V_i\})}$
runs over all the possibilities of circling external points and vertices,
excepted the two terms where all or none of them are circled. 

If we remain in the general case of an out-of-equilibrium field theory, the
result of BDN cannot be generalized in totality because the KMS identity is not
true. More precisely, the configurations of Fig.~\ref{cancelled}-(a) can be
removed from the sum, thanks to the identity Eq.~(\ref{rotconstraint1}) which
takes here a very simple form:
\begin{equation}
{\imb \Gamma}^{\{+\cdots+\}}(\{k_i\})=0\; .
\end{equation}
But in the general case the configurations of Fig.~\ref{cancelled}-(b)
cannot. To remove them also, one must restrict to an equilibrium theory, so
that holds the relation Eq.~(\ref{rotKMS}), which takes here the form:
\begin{equation}
\sum\limits_{\{X_i=\pm\}} {\imb \Gamma}^{\{X_i\}}(\{k_i\}) 
\prod\limits_{\{i|X_i=+\}} b(k^0)b(-k^0)(1+e^{-\beta k_i^0})
\prod\limits_{\{i|X_i=-\}} (1-e^{-\beta k_i^0})=0\; .
\end{equation}
 If
thermal equilibrium holds, then we have only configurations of circles that
correspond to cuts through the diagram:
\begin{equation}
{\rm Im}\left({\imb G}^{\{X_i\}}\right)(\{k_i\}) = -{1\over 2}
\sum\limits_{\{V_i=+++,++-,\cdots\}} \sum\limits_{\rm cuts} {\imb
G}^{(\{X_i\}\{V_i\})}(\{k_i\})\; .
\label{kelBDN}
\end{equation}

\subsection{The $F/\bar{F}$ basis}
\label{FFbar}
\subsubsection{Cutting rules in the ``$F/\bar{F}$" formalism}

In this formalism, one implements also the KMS relation of Eq.~(\ref{KMS}) in
order to make a second component of the matrix propagator vanish. There are
more than one possibility to achieve that. In one of them, we obtain for the
non vanishing components the usual zero-temperature Feynman propagator and its
complex conjugate. Of course, since the statistical factors have disappeared in
the propagator, the new vertices must contain the information relative to the
temperature.

A transformation enabling this for bosons\footnote{A similar transformation can
be performed for fermionic fields. Since it results in propagators that do not
contain the statistical factors, it leads to the same propagators as in the
bosonic case. The difference between bosons and fermions will then appear only
in the vertices. The details of the $F/\bar{F}$ formalism with fermions can be
found in \cite{EijckKW1}.}
 is generated by the following $U$
matrix:
\begin{Eqnarray}
U(k)=&&\theta(k^0)\pmatrix{&c(k^0) & - e^{-\beta k^0} c(k^0)\cr 
& - d(k^0) &  d(k^0)\cr}\nonumber\\
&&\quad+\theta(-k^0) {{n_{_{B}}(-k^0)e^{-\beta k^0}}\over{c(-k^0)d(-k^0)}} 
\pmatrix{&
d(-k^0) & -d(-k^0)\cr & -e^{\beta k^0} c(-k^0) &  c(-k^0)\cr}\; ,
\label{FFrot}
\end{Eqnarray}
where $c(k^0)$ and $d(k^0)$ are arbitrary 
functions of
the energy, non vanishing at the point $k^0=0$.

Performing this transformation, we obtain:
\begin{eqnarray}
&&{\imb G}^{XY}(k)=\pmatrix{&\Delta_{_{F}}(k)
& 0 \cr
& 0 & \Delta_{_{\bar{F}}}(k) \cr}\;
,\\
&&{\imb G}^{\underline{X}Y}(k)=
\pmatrix{&2\pi\theta(-k^0)\delta(k^2-m^2) 
& 0 \cr
& 0 & 2\pi\theta(k^0)\delta(k^2-m^2) \cr}\; ,\\
&&{\imb G}^{X\underline{Y}}(k)=
\pmatrix{&2\pi\theta(k^0)\delta(k^2-m^2) 
& 0 \cr
& 0 & 2\pi\theta(-k^0)\delta(k^2-m^2) \cr}\; ,\\
&&{\rm and}\nonumber\\
&&{\imb G}^{\underline{XY}}(k)=
\pmatrix{&\Delta_{_{\bar{F}}}(k)
& 0 \cr
& 0 &\Delta_{_{F}}(k)  \cr}\; ,
\end{eqnarray}
where we defined
\begin{eqnarray}
&&\Delta_{_{F}}(k)\equiv i{\cal P}{1\over{k^2-m^2}}+\pi\delta(k^2-m^2)\\
&&\Delta_{_{\bar{F}}}(k)\equiv
-i{\cal P}{1\over{k^2-m^2}}+\pi\delta(k^2-m^2)\; .
\end{eqnarray}

Here again, the difficulty in using this formalism for practical calculations
lies in the complexity of the vertices:
\begin{Eqnarray}
&&{{{\imb g}^{X_1X_2X_3}(k_1,k_2,k_3)}\over{-ig}}\nonumber\\
&&\qquad= \prod\limits_{\{i|X_i=F\}}
\left[\theta(k_i^0)c(k_i^0)+\theta(-k_i^0){{e^{-\beta
k_i^0}n_{_{B}}(-k_i^0)}\over{c(-k_i^0)}}\right] 
\prod\limits_{\{i|X_i=\bar{F}\}}
\left[\theta(k_i^0)d(k_i^0) + \theta(-k_i^0){{n_{_{B}}(-k_i^0)}\over{d(-k_i^0)}}
\right]\nonumber\\
&&\quad- \prod\limits_{\{i|X_i=F\}} 
\left[\theta(k_i^0)c(k_i^0)+\theta(-k_i^0)
{{n_{_{B}}(-k_i^0)}\over{c(-k_i^0)}}\right]
\prod\limits_{\{i|X_i=\bar{F}\}}
\left[\theta(k_i^0)e^{\beta k_i^0}d(k_i^0) + 
\theta(-k_i^0){{n_{_{B}}(-k_i^0)}\over{d(-k_i^0)}}
\right] \; .
\end{Eqnarray}
In particular, there is no simple choice of the functions $c(k^0)$ and $d(k^0)$
enabling one to simplify significantly these vertices. 
In this particular formalism, the constraints provided by the relations
Eq.~(\ref{rotconstraint1}) and Eq.~(\ref{rotKMS}) are:
\begin{equation}
\sum\limits_{\{X_i=F,\bar{F}\}} {\imb \Gamma}^{\{X_i\}}(\{k_i\})
\prod\limits_{\{i|X_i=F\}} \theta(-k_i^0) c(-k_i^0) (1-e^{\beta k_i^0})
\prod\limits_{\{i|X_i=\bar{F}\}}
{{\theta(k_i^0)}\over{d(k_i^0)}} =0\; ,
\end{equation}
and
\begin{equation}
\sum\limits_{\{X_i=F,\bar{F}\}} {\imb \Gamma}^{\{X_i\}}(\{k_i\})
 \prod\limits_{\{i|X_i=F\}} {{\theta(k_i^0)}\over{c(k_i^0)}}
  \prod\limits_{\{i|X_i=\bar{F}\}}
\theta(-k_i^0) d(-k_i^0) (e^{-\beta k_i^0}-1)=0\; .
\label{FFKMS}
\end{equation}
If we are at thermal equilibrium, both of them are true, and the imaginary part
of a ``$F/\bar{F}$" Green's function is given by:
\begin{equation}
{\rm Im}\left({\imb G}^{\{X_i\}}\right)(\{k_i\}) = -{1\over 2}
\sum\limits_{\{V_i=FFF,FF\bar{F},\cdots\}} \sum\limits_{\rm cuts} {\imb
G}^{(\{X_i\}\{V_i\})}(\{k_i\})\; .
\end{equation}

\subsection{Non-equilibrium ``$F/\bar{F}$" formalism}

Since the derivation of the propagators of this formalism makes an explicit use
of the KMS relation, the ``$F/\bar{F}$" cannot be generalized to non
equilibrium situations without changes. For example, the propagators we get by
the transformation Eq.~(\ref{FFrot}) without using the KMS relation
are\footnote{Contrary to the case of thermal equilibrium, the
out-of-equilibrium propagators depend on the statistics. We
reproduce here only the result for bosons. The case of fermions has been
derived also, and differs from the bosonic one by the usual changes:
$n_{_{B}}(k^0)\to n_{_{F}}(k^0)$ and $G^{12}(k)-e^{-\beta k^0}G^{21}(k)\to
G^{12}(k)+e^{-\beta k^0}G^{21}(k)$, plus a global minus sign for some
components.}:
\begin{eqnarray}
&&\!\!\!\!\!\!\!\!\!\!\!\!{\imb G}^{XY}(k)= \pmatrix{
&G^{11}(k)-\theta(k^0)G^{12}(k)-\theta(-k^0)G^{21}(k)\!\!\!\!\!\!\!\!\!\!\!\!
& \theta(k^0)\Theta(k)
\displaystyle{{c(k^0)}\over{d(k^0)}}\cr
&\theta(-k^0)\Theta(-k)
\displaystyle{{c(-k^0)}\over{d(-k^0)}}
&\!\!\!\!\!\!\!\!\!\!\!\!
\theta(k^0)G^{21}(k)+\theta(-k^0)G^{12}(k)-G^{11}(k)\cr}\; ,\\
&&\!\!\!\!\!\!\!\!\!\!\!\!{\imb G}^{\underline{X}Y}(k)= \pmatrix{
&\theta(-k^0)(G^{12}(k)-G^{21}(k))
&\theta(k^0)\Theta(k)
\displaystyle{{c(k^0)}\over{d(k^0)}}\cr
&\theta(-k^0)\Theta(-k)
\displaystyle{{c(-k^0)}\over{d(-k^0)}}
&\theta(k^0)(G^{21}(k)-G^{12}(k))\cr}\; ,\\
&&\!\!\!\!\!\!\!\!\!\!\!\!{\imb G}^{X\underline{Y}}(k)= \pmatrix{
&\theta(k^0)(G^{21}(k)-G^{12}(k))
&\theta(k^0)\Theta(k)
\displaystyle{{c(k^0)}\over{d(k^0)}}\cr
&\theta(-k^0)\Theta(-k)
\displaystyle{{c(-k^0)}\over{d(-k^0)}}
&\theta(-k^0)(G^{12}(k)-G^{21}(k))\cr}\; ,\\
&&\!\!\!\!\!\!\!\!\!\!\!\!{\rm and}\nonumber\\
&&\!\!\!\!\!\!\!\!\!\!\!\!{\imb G}^{\underline{XY}}(k)= \pmatrix{
&\theta(k^0)G^{21}(k)+\theta(-k^0)G^{12}(k)-G^{11}(k)\!\!\!\!\!\!\!\!\!\!\!\!
&\theta(k^0)\Theta(k)
\displaystyle{{c(k^0)}\over{d(k^0)}}\cr
&\theta(-k^0)\Theta(-k)
\displaystyle{{c(-k^0)}\over{d(-k^0)}}
&\!\!\!\!\!\!\!\!\!\!\!\!
G^{11}(k)-\theta(k^0)G^{12}(k)-\theta(-k^0)G^{21}(k)\cr}\; ,
\end{eqnarray}
where we denoted
\begin{equation}
\Theta(k)\equiv G^{12}(k)-e^{-\beta k^0}G^{21}(k)\; .
\end{equation}
This quantity, vanishing at thermal equilibrium, quantifies how far we are from
the equilibrium at the temperature $T$. 
We notice that the non-diagonal components are now non-vanishing (but remain
small if we are close to equilibrium at temperature $T$), whereas the vertices
remain the same, so that this formalism is not adapted to non equilibrium
calculations. The reason why both of the non-diagonal components are non zero
whereas we still use the first constraint Eq.~(\ref{constraint1}) lies in the
fact that the ``$F/\bar{F}$" formalism does not implement the two constraints
Eq.~(\ref{constraint1}) and Eq.~(\ref{KMS}) individually, but rather implements
two distinct linear combinations of these identities: therefore, both of these
linear combinations are non zero if the KMS relation is not true anymore.
Moreover, if we are out of equilibrium, the KMS relation Eq.~(\ref{FFKMS}) is
not satisfied, and we must keep all the terms that contain
configurations of circles of the type of Fig.~\ref{cancelled}-(b).

\subsection{The $R/A$ basis}
\label{RA}
\subsubsection{Cutting rules in the ``R/A" formalism}

The ``R/A" formalism is of the same kind as the ``$F/\bar{F}$" one since it
implements the two constraints Eq.~(\ref{constraint1}) and Eq.~(\ref{KMS}). It
has therefore two vanishing components in its propagator. The difference lies
in the nature of the non vanishing components, which are now the zero
temperature retarded and advanced propagators. A transformation matrix that
leads to that formalism for bosons is 
(the remarks made for the $F/\bar{F}$ formalism
concerning fermions are also valid in this paragraph):
\begin{equation}
U(k)={1\over{a(-k^0)}} \pmatrix{
&a(k^0)a(-k^0)
&-a(k^0)a(-k^0)\cr
&- n_{_{B}}(-k^0)
&- n_{_{B}}(k^0)\cr}\; ,
\end{equation}
where $a(k^0)$ is an arbitrary function of $k^0$, non vanishing at $k^0=0$.

The propagators we obtain by this transformation are:
\begin{eqnarray}
&&{\imb G}^{XY}(k)= \pmatrix{
&0 & \Delta_{_{A}}(k)\cr
&\Delta_{_{R}}(k) &0\cr}\; ,\\
&&{\imb G}^{\underline{X}Y}(k)= \pmatrix{
&0 & \Delta_{_{A}}(k)-\Delta_{_{R}}(k)\cr
&0 &0\cr}\; ,\\
&&{\imb G}^{X\underline{Y}}(k)= \pmatrix{
&0 & 0\cr
&\Delta_{_{R}}(k)-\Delta_{_{A}}(k) &0\cr}\; ,\\
&&{\rm and}\nonumber\\
&&{\imb G}^{\underline{XY}}(k)= \pmatrix{
&0 & -\Delta_{_{R}}(k)\cr
&-\Delta_{_{A}}(k) &0\cr}\; .
\end{eqnarray}
As usual for such a transformation, the thermal information is now carried by
the vertices:
\begin{equation}
{{{\imb g}^{X_1X_2X_3}(k_1,k_2,k_3)}\over{-ig}}=
{{\prod\limits_{\{i|X_i=A\}}a(k_i^0)}\over{\prod\limits_{\{i|X_i=R\}}
a(-k_i^0)}} \prod\limits_{\{i|X_i=R\}} n_{_{B}}(k_i^0)
\left[\prod\limits_{\{i|X_i=R\}} e^{\beta k_i^0}-1
\right]\; .
\end{equation}
To precise the vertices, we must make a choice for the function $a(k^0)$. A
convenient choice is $a(k^0)=- n_{_{B}}(k^0)$.  With such a choice, we have 
for a
vertex coupling three bosons:
\begin{eqnarray}
&&{\imb g}^{^{AAA}}={\imb g}^{^{RRR}}=0\\
&&{\imb g}^{^{ARR}}={\imb g}^{^{RAR}}={\imb g}^{^{RRA}}=-ig\\
&&{\imb g}^{^{RAA}}(k_1,k_2,k_3)=ig(1+n_{_{B}}(k_2^0)+n_{_{B}}(k_3^0))\\
&&{\imb g}^{^{ARA}}(k_1,k_2,k_3)=ig(1+n_{_{B}}(k_1^0)+n_{_{B}}(k_3^0))\\
&&{\imb g}^{^{AAR}}(k_1,k_2,k_3)=ig(1+n_{_{B}}(k_1^0)+n_{_{B}}(k_2^0))\; .
\end{eqnarray}
 We note that the
vertices with only advanced or only retarded labels are vanishing. This
property is in fact general for any $n$-point Green's function in the ``R/A"
formalism, since the constraints Eq.~(\ref{rotconstraint1}) and
Eq.~(\ref{rotKMS}) are now:
\begin{eqnarray}
&&{\imb \Gamma}^{\{R\cdots R\}}(\{k_i\})=0\\
&&{\imb \Gamma}^{\{A\cdots A\}}(\{k_i\})=0\; .
\end{eqnarray}
Thanks to these relations, the configurations of circles depicted in
Fig.~\ref{cancelled} both cancel in the calculation of the imaginary part, so
that we are left with:
\begin{equation}
{\rm Im}\left({\imb G}^{\{X_i\}}\right)(\{k_i\}) = -{1\over 2}
\sum\limits_{\{V_i=RRA,RAR,\cdots\}} \sum\limits_{\rm cuts} {\imb
G}^{(\{X_i\}\{V_i\})}(\{k_i\})\; .
\end{equation}
This formalism appears to be particularly simple for the application of cutting
rules. Indeed, each matrix propagator have at least two vanishing components,
and the propagators connecting circled vertices to uncircled ones even have
three zero components. The presence of these zeros has the property to reduce
considerably the number of terms left by the sum over the type of the internal
vertices. Moreover, contrary to the ``$F/\bar{F}$" formalism, the vertices
remain rather simple.

\subsubsection{Non-equilibrium ``R/A" formalism}

Again, some of the nice features of this formalism disappear when one does not
use the KMS relation any longer. The propagators we are left with in such
circumstances are:
\begin{eqnarray}
&&{\imb G}^{XY}(k)= \pmatrix{&0 & G^{11}(k)-G^{21}(k)\cr
&G^{11}(k)-G^{12}(k) 
&\displaystyle{{e^{\beta k^0}\Theta(k)}
\over{a(k^0)a(-k^0)(e^{\beta k^0}-1)}}\cr}\; ,\\
&&{\imb G}^{\underline{X}Y}(k)= \pmatrix{&0 &G^{12}(k)-G^{21}(k) \cr
&0 
&\displaystyle{{e^{\beta k^0}\Theta(k)}
\over{a(k^0)a(-k^0)(e^{\beta k^0}-1)}}\cr}\; ,\\
&&{\imb G}^{X\underline{Y}}(k)= \pmatrix{&0 & 0\cr
&G^{21}(k)-G^{12}(k) 
&\displaystyle{{e^{\beta k^0}\Theta(k)}
\over{a(k^0)a(-k^0)(e^{\beta k^0}-1)}}\cr}\; ,\\
&&{\rm and}\nonumber\\
&&{\imb G}^{\underline{XY}}(k)= \pmatrix{&0 & G^{12}(k)-G^{11}(k)\cr
&G^{21}(k)-G^{11}(k) 
&\displaystyle{{e^{\beta k^0}\Theta(k)}
\over{a(k^0)a(-k^0)(e^{\beta k^0}-1)}}\cr}\; .
\end{eqnarray}
Compared to what happens out of equilibrium in the ``$F/\bar{F}$" basis, we
obtain at least one vanishing component and one component which is small if we
consider a situation close to a thermal equilibrium at temperature $T$. Again,
we have an extra zero component in the propagators connecting circled to
uncircled vertices. Therefore, the ``R/A" formalism appears to be the simplest
one, even out of equilibrium.

\subsection{The case of self-energies}
\label{self}
\subsubsection{Spectral representation of self-energies}

Since the calculation of the imaginary part of two-point proper functions is of
great importance when one is calculating production rates or decay rates of
some particle, we devote a separate paragraph to the study of self-energies,
having in mind the calculation of their imaginary part. Before going on, one
should emphasize that in general, we are not interested in the self-energy
expressed in a particular formalism, but rather in the function that enters the
pole shift of a resummed propagator.
In order to make the connection between this quantity and the
self-energies in various formalisms, it is convenient to start with the
spectral representation of two-point proper functions. 
This implies that we restrict ourselves to equilibrium theories since the KMS
identity is required to derive the spectral representation of Green's
functions.
Let us write it first in
the ``1/2" formalism, where it is well known (see for instance
\cite{Evans5,FetteW1}):
\begin{eqnarray}
&&-i\Sigma^{11}(k_0,{\imb k})=\int\limits_{-\infty}^{+\infty} dE\;
\sigma(E,{\imb k})\, G^{11}(k_0,E)\; ,\\
&&-i\Sigma^{22}(k_0,{\imb k})=\int\limits_{-\infty}^{+\infty} dE\;
\sigma(E,{\imb k})\, G^{22}(k_0,E)\; ,\\
&&-i\Sigma^{12}(k_0,{\imb k})=-\int\limits_{-\infty}^{+\infty} dE\;
\sigma(E,{\imb k})\, G^{12}(k_0,E)\; ,\\
&&-i\Sigma^{21}(k_0,{\imb k})=-\int\limits_{-\infty}^{+\infty} dE\;
\sigma(E,{\imb k})\, G^{21}(k_0,E)\; ,
\end{eqnarray}
where $\sigma(E,{\imb k})$ is a real function, and where $G^{ij}(k_0,E)$
denotes the propagator $G^{ij}$ in which the on-shell energy
$\omega_k\equiv\surd{({\imb k}^2+m^2)}$ has been replaced by $E$. The minus
signs in the $12$ and $21$ components are due to the opposite sign of the type
$2$ vertices. For later convenience, we can summarize these four relations by
the single equation:
\begin{equation}
-i\Sigma^{ij}(k_0,{\imb k})= \int\limits_{-\infty}^{+\infty} dE\;
\sigma(E,{\imb k})\,\sum\limits_{\{k,l=1,2\}} \tau_3^{ik} \tau_3^{jl}
G^{kl}(k_0,E)\; ,
\label{self12}
\end{equation}
where $\tau_3$ is the Pauli matrix along the $z$-direction. Using now the
inverse of Eq.~(\ref{defrot}) and Eq.~(\ref{rotvertex}) in the case of
$2$-point functions, we get:
\begin{equation}
-i{\imb \Sigma}^{XY}(k)=\int\limits_{-\infty}^{+\infty} dE\; \sigma(E,{\imb
k})\,\sum\limits_{\{U,V=\alpha,\bar{\alpha}\}}W^{XU}(k)W^{YV}(-k)
{\imb G}^{UV}(k_0,E)\; ,
\label{selfrot}
\end{equation}
where we defined
\begin{equation}
W^{XU}(k)\equiv\sum\limits_{\{i,k=1,2\}} V^{Xi}(k)\tau_3^{ik}
\left(U^{-1}\right)^{kU}(k)\; .
\end{equation}

For the Keldysh formalism, we obtain
\begin{equation}
W(k)=\pmatrix{&0&1\cr&1&0\cr}\; ,
\end{equation}
so that we have
\begin{eqnarray}
&&-i{\imb \Sigma}^{++}(k_0,{\imb k})=\int\limits_{-\infty}^{+\infty} dE\;\sigma(E,{\imb
k})\,{\imb G}^{--}(k_0,E)=0\; ,\\
&&-i{\imb \Sigma}^{--}(k_0,{\imb k})=\int\limits_{-\infty}^{+\infty} dE\;\sigma(E,{\imb
k})\,{\imb G}^{++}(k_0,E)\; ,\\
&&-i{\imb \Sigma}^{+-}(k_0,{\imb k})=\int\limits_{-\infty}^{+\infty} dE\;\sigma(E,{\imb
k})\,{\imb G}^{-+}(k_0,E)\; ,\\
&&{\rm and}\nonumber\\
&&-i{\imb \Sigma}^{-+}(k_0,{\imb k})=\int\limits_{-\infty}^{+\infty} dE\;\sigma(E,{\imb
k})\,{\imb G}^{+-}(k_0,E)\; .
\end{eqnarray}

In the case of the ``$F/\bar{F}$" formalism, we get:
\begin{equation}
W(k)=\pmatrix{&1&0\cr&0&-1\cr}\; ,
\end{equation}
so that we have
\begin{eqnarray}
&&-i{\imb \Sigma}^{^{FF}}(k_0,{\imb k})=\int\limits_{-\infty}^{+\infty} dE\;\sigma(E,{\imb
k})\,{\imb G}^{^{FF}}(k_0,E)\; ,\\
&&-i{\imb \Sigma}^{^{\bar{F}\bar{F}}}(k_0,{\imb k})=
\int\limits_{-\infty}^{+\infty} dE\;\sigma(E,{\imb
k})\,{\imb G}^{^{\bar{F}\bar{F}}}(k_0,E)\; ,\\
&&-i{\imb \Sigma}^{^{F\bar{F}}}(k_0,{\imb k})=
-\int\limits_{-\infty}^{+\infty} dE\;\sigma(E,{\imb
k})\,{\imb G}^{^{F\bar{F}}}(k_0,E)=0\; ,\\
&&{\rm and}\nonumber\\
&&-i{\imb \Sigma}^{^{\bar{F}F}}(k_0,{\imb k})=
-\int\limits_{-\infty}^{+\infty} dE\;\sigma(E,{\imb
k})\,{\imb G}^{^{\bar{F}F}}(k_0,E)=0\; .
\end{eqnarray}

Finally, for the ``R/A" formalism, we have:
\begin{equation}
W(k)=\pmatrix{&0&1\cr&1&0\cr}\; ,
\end{equation}
which leads to
\begin{eqnarray}
&&-i{\imb \Sigma}^{^{RR}}(k_0,{\imb k})=\int\limits_{-\infty}^{+\infty} dE\;\sigma(E,{\imb
k})\,{\imb G}^{^{AA}}(k_0,E)=0\; ,\\
&&-i{\imb \Sigma}^{^{AA}}(k_0,{\imb k})=
\int\limits_{-\infty}^{+\infty} dE\;\sigma(E,{\imb
k})\,{\imb G}^{^{RR}}(k_0,E)=0\; ,\\
&&-i{\imb \Sigma}^{^{RA}}(k_0,{\imb k})=
\int\limits_{-\infty}^{+\infty} dE\;\sigma(E,{\imb
k})\,{\imb G}^{^{AR}}(k_0,E)\; ,\\
&&{\rm and}\nonumber\\
&&-i{\imb \Sigma}^{^{AR}}(k_0,{\imb k})=
\int\limits_{-\infty}^{+\infty} dE\;\sigma(E,{\imb
k})\,{\imb G}^{^{RA}}(k_0,E)\; .
\end{eqnarray}

\subsubsection{Imaginary part of self-energies in various formalisms}

Then, thanks to these relations and to the fact that the spectral function
$\sigma(E,{\imb k})$ is the same in all the formalisms, we can relate the
imaginary part of self-energies in various formalisms and choose the formalism
that leads to the simplest calculations. In particular, we obtain for the
imaginary part of the pole shift of the Feynman propagator:
\begin{eqnarray}
{\rm Im}\left({\imb \Sigma}^{^{FF}}(k)\right)&&
=\pi\int\limits_{-\infty}^{+\infty} dE\;
\sigma(E,{\imb k})\, \delta(k_0^2-E^2)\\
&&={\rm Im}\left({\imb \Sigma}^{^{\bar{F}\bar{F}}}(k)\right)\\
&&={{-i{\imb
\Sigma}^{--}(k)}\over{2b(k_0)b(-k_0)(1+2n_{_{B}}(|k_0|))}} \\
&&=\epsilon(k_0){\rm Im}\left({\imb \Sigma}^{-+}(k)\right)=\epsilon(-k_0)
{\rm Im}\left({\imb \Sigma}^{+-}(k)\right)\\
&&=\epsilon(k_0){\rm Im}\left({\imb \Sigma}^{^{RA}}(k)\right)
=\epsilon(-k_0){\rm Im}\left({\imb \Sigma}^{^{AR}}(k)\right)\\
&&={{{\rm Im}\left(\Sigma^{11}(k)\right)}\over{1+2n_{_{B}}(|k_0|)}}
={{{\rm Im}\left(\Sigma^{22}(k)\right)}\over{1+2n_{_{B}}(|k_0|)}}\\
&&={{i\Sigma^{12}(k)}\over{2(\theta(-k_0)+n_{_{B}}(|k_0|))}}=
{{i\Sigma^{21}(k)}\over{2(\theta(k_0)+n_{_{B}}(|k_0|))}}\; .
\end{eqnarray}

\subsubsection{Calculation of self-energies in the ``R/A" formalism}

Practically, it appears that the ``R/A" formalism leads to very compact
calculations for the imaginary part of self-energies. In particular, looking at
the place of the vanishing terms in the ``R/A" matrix propagators, and using
the concept of ``$\epsilon$-flow" introduced in \cite{Gueri2}, we can prove the
following statements:
\begin{itemize}
\item[(i)] The connected set of uncircled vertices must be linked to the
external leg on which enters the retarded momentum.

\item[(ii)] The sum over the internal indices $R$ or $A$ contains a single non
vanishing term if the cut goes through all the loops of the diagram ({\it i.e.}
the configuration of the internal $R$ or $A$ indices is totally constrained by
such a cut).

\item[(iii)] If the cut is more general and keeps some loops uncut, it divides
the diagram in two amplitudes, one of them being of the $AR\cdots R$ type while
the other one is the complex conjugate of a $RA\cdots A$ amplitude. This is
another proof of a result contained in \cite{Gueri2}.  Moreover, these
amplitudes are quite simple analytic continuations from the imaginary time
formalism. Nevertheless, as in the ``1/2" cutting rules, the interpretation of
the imaginary part as a sum of products of amplitudes is not completely
meaningful since each of these products is not defined by itself due to
pathologic products of distributions.

\end{itemize}

To conclude this paragraph, one can say that the ``R/A" formalism appears to be
quite simple to calculate imaginary parts of self-energies, because of the
presence of many vanishing components in the propagators of this formalism.
Moreover, the fact that it gives expressions in which the statistical weights
are factorized in the vertices proves to be very helpful to clarify the
calculations.  The ``R/A" formalism seems to be the only one for which BDN's
formula is more efficient than the ``1/2" formalism used with the KS rules. For
instance, in the case of the example already considered on Fig.~\ref{KS}, BDN's
formula in association with the ``R/A" formalism leads to three terms, to be
compared with the eight (only four in practice, since they can be combined by
pairs differing only by a simple factor) terms generated by the method of KS.

\section{Summary}
\label{concl}

The first result of this paper is to make a detailed comparison of two
apparently contradictory approaches for the calculation of the imaginary part
of thermal Green's functions in the ``1/2" real time formalism. It appeared
that the two methods are in fact completely equivalent in the sense that they
lead to the same result. Moreover, despite the fact that Bedaque, Das and
Naik's method seems to lead to a
more common interpretation of the imaginary part in terms of products of
amplitudes, this interpretation is partly spoiled by the fact that the terms
that would correspond to such an interpretation are in fact ill-defined when
considered individually. From the point of view of the length of the
calculations, the method of Kobes and Semenoff appeared to be much more
efficient.

The second part of the paper has been devoted to the generalization of the
cutting rules already obtained in the ``1/2" formalism, in order to give
similar rules for other bases of the real time formalism. It appeared that such
rules are in fact quite general, and that the result of Bedaque, Das and Naik
can be generalized to any new formalism, provided we restrict ourselves to
equilibrium field theories. Among these formalisms, the ``retarded/advanced"
one appeared to be quite efficient for the calculation of the imaginary part of
thermal Green's functions, and even more in the case of self-energies.

Each time it was possible, we indicated how the formulas would change if one
were dropping the KMS relation, thereby enabling the possibility of an
out-of-equilibrium system. Here again, the ``R/A" formalism seems to be the
simplest one. Nevertheless, it should be emphazised that, at present date, the
simple generalization of equilibrium Feynman rules to a non-equilibrium
situation based on the replacement of the Bose-Einstein statistical weight by
an arbitrary function is not consistent, since the cancelation of the
pathologic products of distributions does not work anymore.

\section{Acknowledgments}

I would like to thank F.~Gu\'erin, M.~le~Bellac and J.~Orloff for discussions.
Moreover, I should thank P.~Aurenche and R.~Kobes for many discussions and for
their careful reading of early versions of this manuscript.

\section{Appendix: $T=0$ limit of the KMS relations}
The purpose of this appendix is to derive the $T\to 0^+$ limit of the KMS
identity for $n$-point vertex functions (Eq.~(\ref{KMS})). Let us first note
that we have
\begin{equation}
\lim_{T\to 0^+}\left[\prod\limits_{\{i|c_i=2\}}e^{-\beta k_i^0}\right]=
\left\{\vcenter{
\hbox{$\;0\quad{\rm if}\quad \sum\limits_{\{i|c_i=2\}}k_i^0 >0$}
\hbox{\quad}
\hbox{$\;+\infty\quad{\rm if}
\quad \sum\limits_{\{i|c_i=2\}}k_i^0 <0\; ,$}}\right.
\end{equation}
and more generally
\begin{equation}
\lim_{T\to 0^+}\left[{{\prod\limits_{\{i|c_i=2\}}e^{-\beta k_i^0}}\over
{\prod\limits_{\{i|c'_i=2\}}e^{-\beta k_i^0}}}\right]=
\left\{\vcenter{
\hbox{$\;0\quad{\rm if}\quad \sum\limits_{\{i|c_i=2\}}k_i^0 >
\sum\limits_{\{i|c'_i=2\}}k_i^0$}
\hbox{\quad}
\hbox{$\;+\infty\quad{\rm if}
\quad \sum\limits_{\{i|c_i=2\}}k_i^0 < 
\sum\limits_{\{i|c'_i=2\}}k_i^0\; .$}}\right.
\end{equation}
The next step is to sort the various sets $\{c_i\}$ of indices according to the
corresponding value of the sum $\sum_{\{i|c_i=2\}}k_i^0$, beginning by the
smallest one. Let us assume that the set $\{c_i\}$ is the one corresponding to
the smallest value of this sum and that this value is negative. Therefore, the
quantity $\prod_{\{i|c_i=2\}}e^{-\beta k_i^0}$ goes to $+\infty$ when $T$
goes to zero, and moreover this term is the one that dominates the sum in the
left hand side of Eq.~(\ref{KMS}). The consequence of this is that the
corresponding $\lim_{T\to 0^+}\Gamma^{\{c_i\}}(\{k_i\})$ should vanish in order
to have a zero right hand side in Eq.~(\ref{KMS}). Then, we consider the set
$\{c'_i\}$ which gives the smallest value just above the one given by
$\{c_i\}$, and if $\sum_{\{i|c'_i=2\}}k_i^0$ is still negative, then the
corresponding $\Gamma^{\{c'_i\}}$ is vanishing in the zero temperature limit.

By repeating these steps until we have considered all the sets $\{c_i\}$
giving a negative value of the sum $\sum_{\{i|c_i=2\}}k_i^0$, we prove the
announced relation:
\begin{equation}
\lim_{T\to 0^{+}} \Gamma^{\{c_i\}}(\{k_i\})\propto
\theta\left(\sum\limits_{\{i|c_i=2\}} k_i^0\right)\; .
\end{equation}

\par\section{Figure Captions}
\begin{figure}
\vskip 1cm
\begin{center}
\leavevmode
\epsfbox{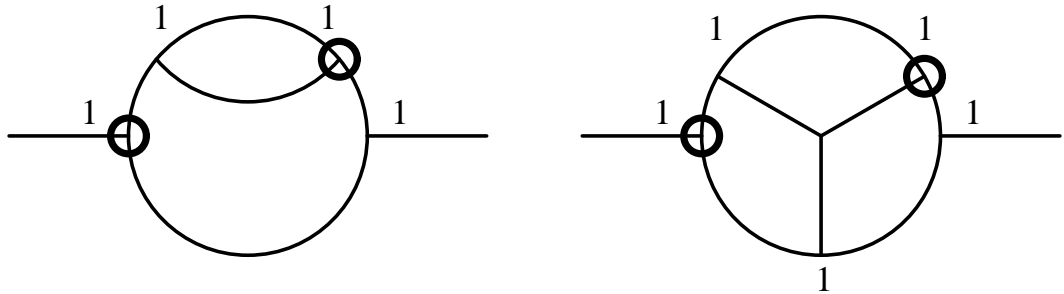}
\end{center}
\caption{Examples of contributions which cannot be
interpreted as cut diagrams in the approach of Kobes and Semenoff.}
\label{uncut}
\end{figure}

\begin{figure}
\begin{center}
\leavevmode
\epsfbox{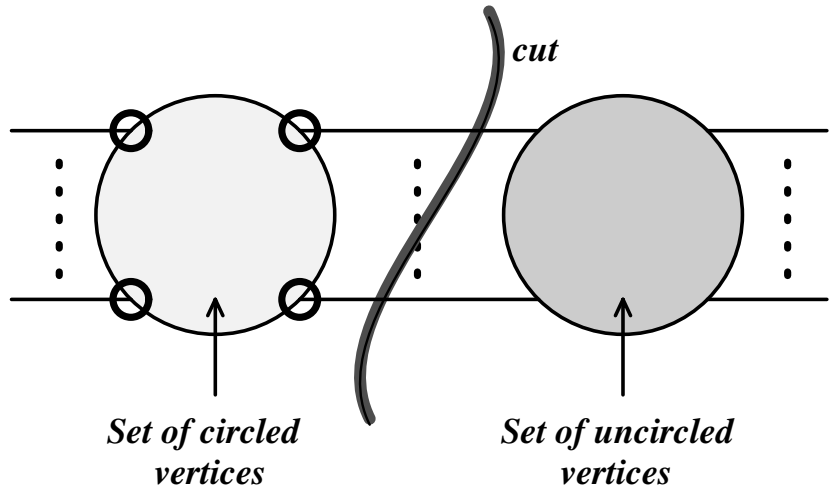}
\end{center}
\caption{Structure of a cut diagram in the approach 
of Bedaque, Das and Naik.}
\label{cut}
\end{figure}

\begin{figure}
\begin{center}
\leavevmode
\epsfbox{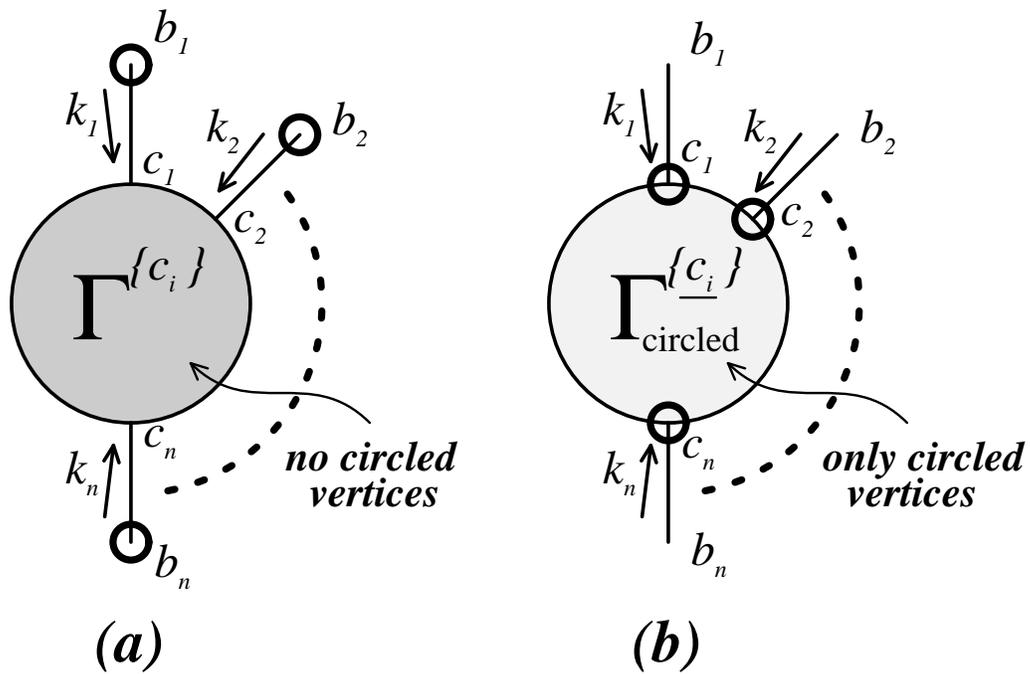}
\end{center}
\caption{Terms that are canceled by the summation
over the type $1$ or $2$ of the internal vertices.}
\label{cancelled}
\end{figure}

\begin{figure}
\begin{center}
\leavevmode
\epsfbox{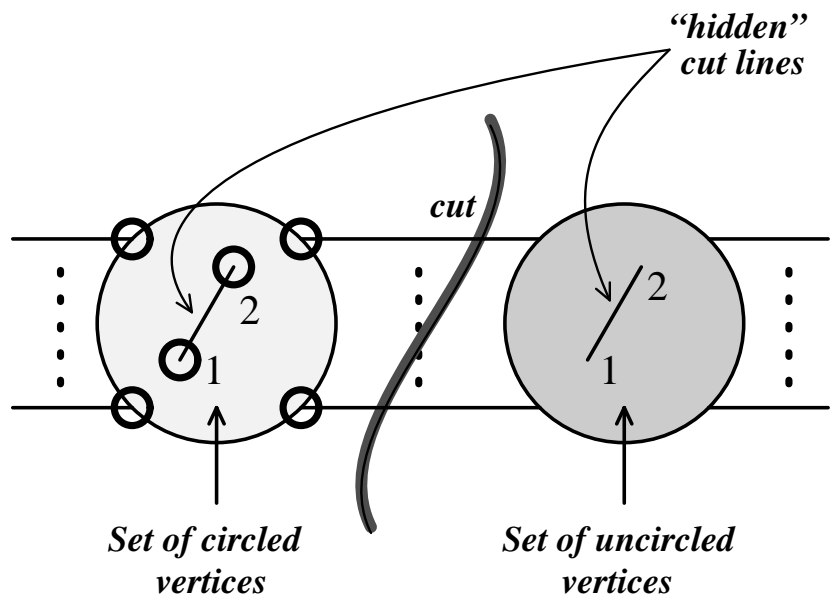}
\end{center}
\caption{``Hidden" cuts in the terms generated in the
approach of Bedaque, Das and Naik.}
\label{hidden}
\end{figure}

\begin{figure}
\begin{center}
\leavevmode
\epsfbox{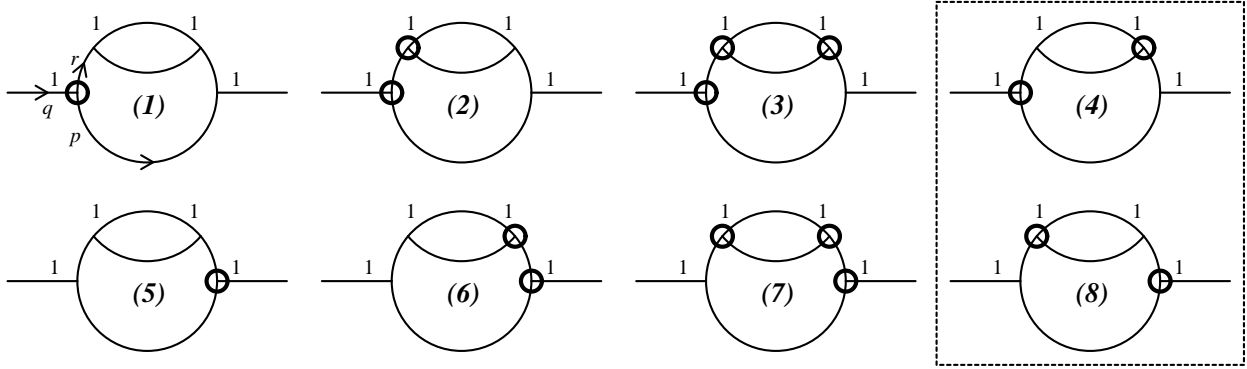}
\end{center}
\caption{Terms generated by the method of Kobes and
Semenoff in the example of Fig.~\ref{uncut}. The two terms in the box 
cannot be
interpreted as contributions of cut diagrams.}
\label{KS}
\end{figure}

\begin{figure}
\begin{center}
\leavevmode
\epsfbox{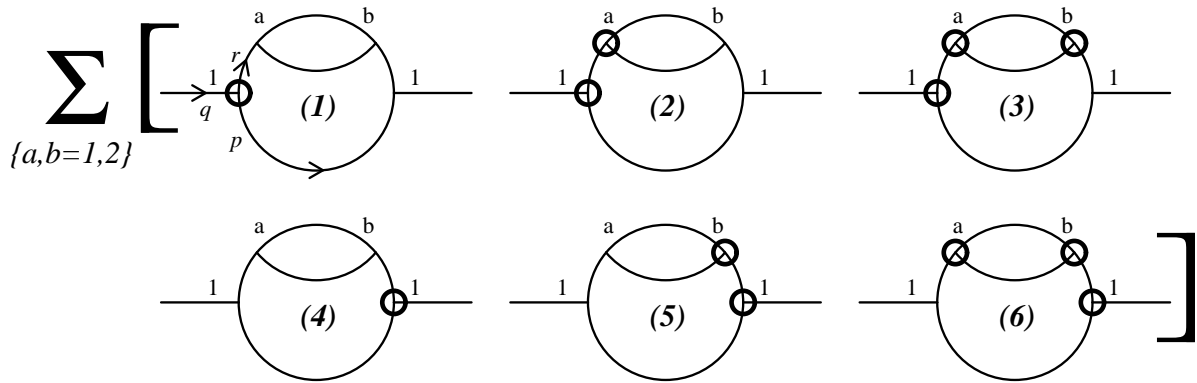}
\end{center}
\caption{Terms generated by the method of Bedaque, Das and Naik
 in the example of Fig.~\ref{uncut}.}
\label{BDN}
\end{figure}

\begin{itemize}
\item[Fig.~\ref{uncut}] Examples of contributions which cannot be
interpreted as cut diagrams in the approach of Kobes and Semenoff.
\item[Fig.~\ref{cut}] Structure of a cut diagram in the approach 
of Bedaque, Das and Naik.
\item[Fig.~\ref{cancelled}] Terms that are canceled by the summation
over the type $1$ or $2$ of the internal vertices.
\item[Fig.~\ref{hidden}] ``Hidden" cuts in the terms generated in the
approach of Bedaque, Das and Naik.
\item[Fig.~\ref{KS}] Terms generated by the method of Kobes and
Semenoff in the example of Fig.~\ref{uncut}. The two terms in the box 
cannot be
interpreted as contributions of cut diagrams.
\item[Fig.~\ref{BDN}] Terms generated by the method of Bedaque, Das
and Naik in the example of  Fig.~\ref{uncut}.
\end{itemize}
\clearpage

\clearpage

\end{document}